\documentclass[epj]{svjour}
 \usepackage{graphicx,amsmath,amssymb}
 \usepackage{axodraw}

\newcommand\epjc[3]  {{Eur.\ Phys.\ J. }{\bf C #1} (#2) #3}

\newcommand\nature[3]{{Nature }{\bf #1} (#2) #3}

\newcommand\pla[3]   {{Phys.\ Lett.\ }{\bf A #1} (#2) #3}
\newcommand\plb[3]   {{Phys.\ Lett.\ }{\bf B #1} (#2) #3}
\newcommand\joupr[3]    {{Phys.\ Rev.\ }{\bf #1} (#2) #3}

\newcommand\jouprb[3]   {{Phys.\ Rev.\ }{\bf B #1} (#2) #3}

\newcommand\jouprd[3]   {{Phys.\ Rev.\ }{\bf D #1} (#2) #3}

\newcommand\prep[3]  {{Phys.\ Rept.\ }{\bf #1} (#2) #3}
\newcommand\jouprl[3]   {{Phys.\ Rev.\ Lett.\ }{\bf #1} (#2) #3}

\newcommand\jourmp[3]   {{Rev.\ Mod.\ Phys.\ }{\bf #1} (#2) #3}

\newcommand{\newjournal}[5]{{#1 }{\bf #3} (#4) #5}

\newcommand\ibid[3]{{ibid.\ }{\bf #1} (#2) #3}

\newcommand{\hepph}[1]{{hep-ph/#1}}

\begin{document}

 \title{Current conservation and ratio rules in magnetic metals with 
Coulomb repulsion}

 \author{Kosuke~Odagiri}

 \institute{
  Electronics and Photonics Research Institute,
  National Institute of Advanced Industrial Science and Technology,
  Tsukuba Central 2,
  1--1--1 Umezono, Tsukuba, Ibaraki 305--8568, Japan
  }

 \abstract{
  From general considerations of spin-symmetry breaking associated with 
(anti-)ferromagnetism in metallic systems with Coulomb repulsion, we 
obtain interesting and simple all-order rules involving the ratios of 
the densities of states. These are exact for ferromagnetism under 
reasonable conditions, and nearly exact for anti-ferromagnetism.
  In the case of ferromagnetism, the comparison with the available 
experimental and theoretical numbers yields favourable results.
 }

 \date{December 2011}

 \maketitle

 \tableofcontents

 \section{Introduction}
 \label{sec_introduction}

 \subsection{Theoretical background}

  As is obvious and well known \cite{leutwyler}, magnetic order breaks 
SU(2) spin symmetry (hereafter called SU(2)$_\mathrm{spin}$). This gives 
rise to gapless excitations in the form of Nambu--Goldstone modes which 
are magnons and, according to the Goldstone theorem, excitations with 
energy gap, which may be called the Higgs bosons.

  These excitations behave like elementary fields, and their interaction 
is central to spin-current conservation but, at the same time, they 
comprise of electrons with which they interact: i.e., they are 
composite. This imposes severe constraints on the properties of these 
fields, which we aim to discuss and exploit in this paper.

  The same situation, of Goldstone fields (i.e., both Goldstone and 
Higgs fields) that are themselves composite objects, arises notably in 
two problems in the context of high-energy physics.
  The first problem is that of axial symmetry breaking at low energy 
scales due to the SU(3)$_C$ strong interaction.
  The Goldstone bosons here are pions.
  The second problem is that of electro-weak symmetry breaking.
  Although in the Standard Model, the Higgs field, or the doublet 
order-parameter field for SU(2)$_L\otimes$U(1)$_Y$, is an elementary 
scalar field, it is an attractive possibility that this field is 
composite and, for example, is composed of quarks which bind strongly at 
high scales due to some interaction (N.B.\ not by the SU(3)$_C$ strong 
interaction which is weakly interacting at those scales). The low-energy 
phenomenology would then not be dissimilar to that of the Standard 
Model, but with some constraints on quantities such as the Higgs-boson 
mass.

  A new approach to these problems, due to Gribov, have appeared in 
refs.~\cite{gribov,gribovewsb,gribovlectures,yurireview}.
  These involve the idea of super-criticality and a self-consistent 
treatment of the fermion and Goldstone fields in the presence of the 
super-critical interaction.
  We shall make use of, and extend, the methods presented therein, to 
the case of magnetism.
  As for the other approaches to these old problems, see, for example, 
ref.~\cite{njl} for an old approach to the first problem, and 
ref.~\cite{topcondensation} for an overview of the various methods and 
techniques developed to handle the second problem.

 \subsection{Outline of the paper}

  Our work concerns systems of electrons (or holes) which interact under 
a generalized Coulomb exchange (i.e., exchange of a generic gapless 
photon). We consider the system in the spin-symmetry-broken phase that 
arise in ferromagnetism and anti-ferromagnetism.

  Our aim is to obtain exact relations between quantities that 
characterize the spin-symmetry-broken phase using the Dyson--Schwinger 
equations. This is possible because of the presence of the 
Ward--Takahashi identities which arise because of the conservation of 
spin symmetry. It turns out that the form of the Coulomb interaction 
does not affect these relations. The Coulomb interaction does affect, 
for instance, the electronic self-energy, but these are incorporated in 
the relations in a general way.

  Before presenting the full analytical framework, we start with the 
simple case of the discussion of ground-state stability in 
ferromagnetism, in sec.~\ref{sec_ferro_ratio_rule}.
  This gives rise to an exact rule for ferromagnetism which involves the 
electronic densities of states.
  We discuss this case with illustrations and a physical interpretation.

  The full framework, which employ current-conservation techniques, is 
developed in secs.~\ref{sec_spin_current} and \ref{sec_parameters}. In 
sec.~\ref{sec_spin_current}, the interaction is worked out and presented 
in the form of Feynman rules. In sec.~\ref{sec_parameters}, the 
parameters of the interaction is worked out. This gives rise to an exact 
rule for anti-ferromagnetism which involves the electronic densities of 
states, but which involves the bare spin exchange energy.

  The conclusions are stated at the end.

 \section{The ferromagnetic ratio rule}
 \label{sec_ferro_ratio_rule}

  Before we introduce the full framework, let us discuss the stability 
of the ferromagnetic ground state as an illustrative example. We do so 
because this is a relatively simple problem, which does not require the 
full formalism, and which nevertheless leads to a strikingly simple and 
useful ratio rule.
  We shall expand the methods introduced here to build the formalism 
later.

 \subsection{Description of the system}

  Let us write the electrons in the SU(2)$_\mathrm{spin}$ doublet form:
 \begin{equation}
  \psi_a\equiv\left(\begin{array}{c}\psi_\uparrow\\\psi_\downarrow
  \end{array}\right).
 \end{equation}

  The Lagrangian has the form:
 \begin{equation}
  \mathcal{L}=\delta_{ab}\psi^*_a\left(
  i\frac{\partial}{\partial t}-\epsilon(-i\nabla)-e\Gamma^\mu A_\mu
  \right)\psi_b
  +\left(\mbox{photon K.E.}\right)
  \label{eqn_Coulombic_Lagrangian}
 \end{equation}
  $A_\mu$ is the electro-magnetic field, of which we shall retain only 
the electrostatic term $A_0$ later, as the contribution of the 3-vector 
potential is suppressed by the speed of light.
  The photon kinetic energy term may then be taken to be $-(\nabla 
A_0)^2/2$, which leads to an electrostatic $1/r$ interaction.
  $\epsilon$ refers to the dispersion relation of the electron. $e$ is 
the electro-magnetic charge, which is defined to be negative for 
electrons. $\Gamma^\mu$ is the vertex function, whose time component is 
$1$ for a Lagrangian of this form.

  In the absence of magnetic order, the system is invariant under both 
the electromagnetic U(1)$_\mathrm{EM}$ and the SU(2)$_\mathrm{spin}$ 
rotations of $\psi$, where the latter is represented by
 \begin{equation}
  \mathcal{U}(\phi_i)\equiv\exp\left(i\sigma^i\phi_i/2\right).
 \end{equation}
  $\sigma^i$ ($i=1,2,3$) are the Pauli matrices and $\phi_i$ are the 
rotation angles. Note that we are referring to global rotations here. 
Invariance under local rotations requires the presence of gauge fields 
such as the electromagnetic field, including the vector potential part
and written in a gauge-invariant fashion.

  There are two conserved currents, which are orthogonal.
  The first is the U(1)$_\mathrm{EM}$ current:
 \begin{equation}
  J^\mu_\mathrm{EM}=\psi^*_a\delta_{ab}\Gamma^\mu\psi_b.
 \end{equation}
  $\mu$ refers to the time-space four-vector index (=0,1,2,3).

  The second is the SU(2)$_\mathrm{spin}$ current. This is written as
 \begin{equation}
  J^{\mu,i}_\mathrm{spin}=\psi^*_a\sigma^i_{ab}\Gamma^\mu\psi_b.
 \end{equation}

  Let us indicate the current diagrammatically by a cross. We see that 
current--current mixing, which we denote as 
$\Pi^{\mu\nu}_\mathrm{mixing}$ and whose lowest order term is given by
 \begin{equation}
  \begin{picture}(100,30)(0,0)
   \ArrowArcn(50,-36)(60,120,60)
   \ArrowArcn(50,66)(60,300,240)
   \Line(17,10)(23,20) \Line(23,10)(17,20)
   \Text(20,21)[b]{$\mu$}
   \Text(8,15)[c]{$\sigma^i_{ab}$}
   \Line(77,10)(83,20) \Line(83,10)(77,20)
   \Text(80,21)[b]{$\nu$}
   \Text(92,15)[c]{$\delta_{ab}$}
  \end{picture}
  \label{eqn_current_mixing}
 \end{equation}
  vanishes by symmetry for all $i$ to all perturbative orders, and 
therefore the two currents are orthogonal.

  When there is magnetic order, there arises, locally, a preferred 
orientation of spin, let us say along $\downarrow$, and this breaks the 
SU(2)$_\mathrm{spin}$ symmetry, viz:
 \begin{equation}
  \mathrm{SU(2)}_\mathrm{spin}\longrightarrow\mathrm{U(1)}_z.
  \label{eqn_SU_two_breaking}
 \end{equation}
  As a result, there arises two Goldstone modes whose coupling is 
proportional to linear combinations of $\sigma^{1,2}$, and a Higgs mode 
whose coupling is proportional to $\sigma^3$. The residual symmetry 
U(1)$_z$ refers to the symmetry under rotation by the generator 
$\sigma^3$:
 \begin{equation}
  \mathcal{U}(\phi_3)\equiv\exp\left(i\sigma^3\phi_3/2\right)\equiv
  \mathrm{diag}\left(e^{i\phi_3/2},e^{-i\phi_3/2}\right).
 \end{equation}
  The form of the effective theory will be discussed later.

 \subsection{U(1) current mixing}

  A result, which is almost trivial but possibly not previously 
discussed explicitly, is that after this symmetry breaking, the currents 
are no longer orthogonal.
  Equation (\ref{eqn_current_mixing}) is easily calculated.
  Of particular interest is the $0-0$ component of 
eqn.~(\ref{eqn_current_mixing}) for $i=3$ (i.e., the U(1)$_z$ current: 
$i=1,2$ vanish) at zero external energy and momenta. The vertex function 
$\Gamma^0$ being equal to $1$ when the electrons are Fermi-liquid-like 
for each spin orientation, we obtain
 \begin{equation}
  \Pi^{00}_\mathrm{mix}=\lim_{q\to0}
  \int\frac{d^{d+1}k}{(2\pi)^{d+1}i}G_\uparrow(k)G_\uparrow(q+k)
  -G_\downarrow(k)G_\downarrow(q+k).
 \end{equation}
  Hence
 \begin{equation}
  \Pi^{00}_\mathrm{mix}
  =g_\uparrow(\epsilon_F)-g_\downarrow(\epsilon_F).
  \label{eqn_current_mixing_after_SSB}
 \end{equation}
  Note that $\Pi$ is defined with a negative sign, that is 
$\Pi=-\mathcal{A}$, where $\mathcal{A}$ is the two-point amplitude.
  $G$ are the electron Green's functions.
  $g(\epsilon_F)$ are the densities of states at Fermi energy.

  As is well known, the Dyson--Schwinger all-order corrections to a fish 
diagram such as that indicated in eqn.~(\ref{eqn_current_mixing}) is 
incorporated by replacing the Green's functions by their all-order 
counterparts and one or the other of the vertices (and not both) by 
their all-order counterpart.
  Equation (\ref{eqn_current_mixing_after_SSB}) is an all-order 
expression in the sense that the Green's functions are arbitrary.
  However, the vertex correction needs care. Let us consider the 
all-order correction to the photon vertex. At zero external energy and 
momenta, the time component of the all-order vertex is given by the 
energy derivative of $G^{-1}$, by virtue of the Ward--Takahashi 
identity. It follows that, if the all-order Green's function is given by
 \begin{equation}
  G(E,\mathbf{k})=\frac{Z}{
  E-\epsilon(\mathbf{k})+i0\mathrm{sgn}(\epsilon(\mathbf{k})-\mu)},
 \end{equation}
  as is the case for the Fermi liquid, then the vertex is corrected by 
$Z^{-1}$.
  On the other hand, there is $Z^2$ coming from $G^2$, and so the net 
result is proportional to $Z$. $Z$ being the correct renormalizing 
factor for the density of states, 
eqn.~(\ref{eqn_current_mixing_after_SSB}) is exact.
  It is not difficult to see that a more general form of $G$ also admits 
this property:
 \begin{equation}
  G(E,\mathbf{k})=\frac{Z(\mathbf{k})}{
  f(E-\epsilon(\mathbf{k})+i0\mathrm{sgn}(\epsilon(\mathbf{k})-\mu))},
  \label{eqn_generalized_green_function}
 \end{equation}
  where $f$ is any function, so long as the density of states is 
definable as the integral of $G$. Thus it is not a necessary condition 
that the system is a Fermi liquid.

  Returning to eqn.~(\ref{eqn_current_mixing_after_SSB}), in general, 
$g_\uparrow$ and $g_\downarrow$ are not equal at the Fermi surface, and 
therefore the currents mix.

  It is worth noting here that the current mixing is zero in the case of 
anti-ferromagnetism, because the two sub-lattice contributions are equal 
and opposite.

  Before proceeding, let us calculate the other fish diagrams.
  Both for the EM current and for the spin U(1)$_z$ current, we obtain:
 \begin{equation}
  \Pi^{00}_\mathrm{EM}=\Pi^{00}_z
  =g_\uparrow(\epsilon_F)+g_\downarrow(\epsilon_F).
  \label{eqn_photon_self_energy}
 \end{equation}
  Again, this is an exact result provided that the Green's functions are 
of the form eqn.~(\ref{eqn_generalized_green_function}) and the densities 
of states can be defined as their integrals.
  Note that although we have retained the subscript `EM' to refer to 
electromagnetism, in reality, we are analyzing electrostatics.

 \subsection{Derivation of the ratio rule}

  Let us now consider the stability of the ferromagnetic ground state.
  To do so, a primary condition is the vanishing of the tadpole:
 \begin{equation}
  \begin{picture}(50,30)(0,0)
   \ArrowArcn(35,15)(15,270,-90)
   \GCirc(20,15){3}{0}
   \Text(8,15)[c]{$\sigma^3$}
  \end{picture}
  \label{eqn_higgs_tadpole}
 \end{equation}
  as is required by the condition that there are no terms that are 
linear in the Higgs field in the effective Lagrangian. In other words, 
the first derivative of free energy as a function of the magnetic order 
parameter must vanish when the ground state is stable. We will also need 
to check that the second derivative is positive. This means the term 
which is bilinear in the Higgs field, or the self-energy of the Higgs 
boson, is positive. The Higgs self-energy is the same as $\Pi^{00}_z$ 
calculated earlier on, up to the square of a coupling constant. This 
is necessarily positive.

  Equation (\ref{eqn_higgs_tadpole}) is calculated easily, and we obtain
 \begin{equation}
  \mathcal{A}^\mathrm{tadpole}_z=
  \int\frac{d^{d+1}k}{(2\pi)^{d+1}i}G_\uparrow(k)-G_\downarrow(k)
  =\rho_\uparrow-\rho_\downarrow.
 \end{equation}
  $\rho$ refers to the total density of states of electrons.
  This is an exact expression since higher-order corrections to the 
tadpole are taken into account by making $G$ all-order and vertex to be 
bare.
  This is always negative if $\downarrow$ is the preferred orientation 
of spin.

  It follows that eqn.~(\ref{eqn_higgs_tadpole}) by itself is non-zero. 
However, the currents mix, and we must incorporate the contribution of 
the photon tadpole, multiplied by the Higgs--photon two-point function 
which has the same form as $\Pi^{00}_\mathrm{mixing}$ calculated in 
eqn.~(\ref{eqn_current_mixing_after_SSB}):
 \begin{equation}
  \begin{picture}(150,30)(0,0)
   \ArrowArcn(50,-36)(60,120,60)
   \ArrowArcn(50,66)(60,300,240)
   \GCirc(20,15){3}{0}
   \Text(8,15)[c]{$\sigma^3_{ab}$}
   \Text(83,19)[lb]{$\delta_{ab}$}
   \GCirc(80,15){3}{0}
   \Photon(80,15)(120,15){3}{3.5}
   \ArrowArcn(135,15)(15,270,-90)
   \GCirc(120,15){3}{0}
  \end{picture}
  \label{eqn_photon_tadpole}
 \end{equation}
  Now, to make this equation all-order, we must include the screening 
effect in the photon propagator, and this has the same form as 
$\Pi^{00}_\mathrm{EM}$ calculated in eqn.~(\ref{eqn_photon_self_energy}). 
Altogether, we obtain
 \begin{equation}
  \mathcal{A}^\mathrm{tadpole}_\mathrm{photon\ part}=
  (\rho_\uparrow+\rho_\downarrow)e\times
  \frac{(g_\uparrow(\epsilon_F)-g_\downarrow(\epsilon_F))e}
  {-(g_\uparrow(\epsilon_F)+g_\downarrow(\epsilon_F))e^2}.
  \label{eqn_photon_tadpole_amplitude}
 \end{equation}

  The two contributions must vanish when added together.
  Although the charge $e$ appears here, whether one takes the charge 
carriers to be electrons or holes is a matter of choice, so $e$ can be 
taken as constant. Hence,
 \begin{equation}
  \frac{g_\downarrow(\epsilon_F)-g_\uparrow(\epsilon_F)}
  {g_\downarrow(\epsilon_F)+g_\uparrow(\epsilon_F)}=
  \frac{\rho_\downarrow-\rho_\uparrow}{\rho_\downarrow+\rho_\uparrow}
  \label{eqn_ferromagnetic_ratio_rule}
 \end{equation}
  or,
 \begin{equation}
  \frac{g_\uparrow(\epsilon_F)}{g_\downarrow(\epsilon_F)}
  =\frac{\rho_\uparrow}{\rho_\downarrow}.
  \label{eqn_ferromagnetic_ratio_rule_simple}
 \end{equation}
  This is our ferromagnetic ratio rule.

  In eqn.~(\ref{eqn_ferromagnetic_ratio_rule}), the left-hand side is 
often called the spin polarization $P$, for example in the context of 
tunnel magneto-resistance.
  The right-hand side is the magnetic moment $n_B=m/\mu_B$ divided by 
the number of carriers $n$, i.e.,
 \begin{equation}
  P=\frac{n_B}{n}.
  \label{eqn_ferromagnetic_ratio_rule_alternative}
 \end{equation}
  Note that the definition of $n$ is ambiguous.
  However, it is a measure of the number of electrons or holes that are 
actively involved in the formation of ferromagnetic order.
  As such, one would expect that its order is estimated by the number of 
carriers in the conduction band.
  If so, we obtain a simple rule of the thumb:
 \begin{equation}
  \left\{\begin{array}{ll}
  g_\downarrow>g_\uparrow & \mathrm{(electrons),}\\
  g_\downarrow<g_\uparrow & \mathrm{(holes).}
  \end{array}\right.
 \end{equation}
  That is, the density of states of the majority-spin charge carriers is 
always greater.
  According to this rule, one expects the density of states to be rising 
with energy for electrons and decreasing with energy for holes. This can 
be a useful rule of the thumb to establish whether a system with certain 
given density of states is likely to become a ferromagnet.

 \subsection{A diagrammatic derivation}

  The preceding derivation is formal and, we believe, complete, but it 
may appear baffling at the start that the sum of apparently only two 
contributions is sufficient to give all-order statements about the 
system.

  Another way to derive the same result, in a more diagrammatic fashion, 
is to consider the mixing between screened photon and the Higgs boson. 
Let us denote the mixed states by $\widetilde\gamma$ and $\widetilde h$. 
Then the sum of the tadpoles
 \begin{equation}
  \begin{picture}(75,30)(0,0)
   \Text(5,15)[c]{$\sigma^3_{ab}$}
   \DashLine(10,15)(40,15){5}
   \ArrowArcn(55,15)(15,270,-90)
   \Text(25,20)[b]{$\widetilde\gamma,\widetilde h$}
  \end{picture}
 \end{equation}
  needs to be zero, whereas the sum of the tadpoles
 \begin{equation}
  \begin{picture}(75,30)(0,0)
   \Text(5,15)[c]{$\delta_{ab}$}
   \DashLine(10,15)(40,15){5}
   \ArrowArcn(55,15)(15,270,-90)
   \Text(25,20)[b]{$\widetilde\gamma,\widetilde h$}
  \end{picture}
 \end{equation}
  is non-zero, and gives a constant contribution to the energy levels of 
the states.

  The sum over the former set of tadpoles can be expanded 
diagrammatically in terms of $\gamma$ and h as:
 \begin{equation}
  \begin{picture}(166,30)(0,0)
   \DashLine(10,15)(20,15){5}
   \GCirc(22.5,15){2.5}{1}
   \Text(33,15)[c]{$+$}
   \DashLine(40,15)(50,15){5}
   \GCirc(52.5,15){2.5}{1}
   \Photon(55,15)(65,15){2}{1.5}
   \GCirc(67.5,15){2.5}{1}
   \Text(78,15)[c]{$+$}
   \DashLine(85,15)(95,15){5}
   \GCirc(97.5,15){2.5}{1}
   \Photon(100,15)(110,15){2}{1.5}
   \GCirc(112.5,15){2.5}{1}
   \DashLine(115,15)(125,15){5}
   \GCirc(127.5,15){2.5}{1}
   \Text(138,15)[l]{$+\cdots$}
  \end{picture}.
 \end{equation}
  It will be seen that if the sum of the first two terms is zero, then 
the remaining contributions, which are proportional to the sum of first 
two terms, vanish automatically. Therefore it suffices to calculate the 
sum of the first two terms.

 \subsection{Examples}

  One way to understand 
eqn.~(\ref{eqn_ferromagnetic_ratio_rule_alternative}) is as a definition 
of $n$.
  This number can then be compared with the other estimates of the 
number of carriers such as by the Hall effect and with the nominal 
number of electrons or holes.

  As a first example, in the case of half metals, and in the ideal case, 
the spin polarization would be perfect, i.e., $P=1$. The magnetization 
will also be perfect, and 
eqn.~(\ref{eqn_ferromagnetic_ratio_rule_alternative}) will be satisfied 
so long as we take $n$ to be equal to $n_B$.

  Next, in the case of a Coulomb system whose renormalized dispersion 
relation is given exactly by 
$\epsilon(\mathbf{k})=(\hbar\mathbf{k})^2/2m_e$, 
the density of states is given by:
 \begin{equation}
  g(\epsilon)=\frac1{\pi^2}\sqrt{m_e^3\epsilon/2}, \quad
  \rho=\frac1{3\pi^2}\sqrt{m_e^3\epsilon_F^3/2}.
 \end{equation}
  We have defined $\rho$ as the integral over whole of the occupied 
states.
  Equation (\ref{eqn_ferromagnetic_ratio_rule_simple}) then admits the 
following solutions only:
 \begin{equation}
  g_\uparrow=g_\downarrow,\quad\mbox{or,}\quad g_\uparrow=0.
 \end{equation}
  That is, either the system is a half metal or there is no 
ferromagnetic order. This is consistent with the observation that group 
1 elements are not ferromagnetic, and also with the observation that the 
Fermi gas system with quadratic (bare) dispersion relation is only 
weakly ferromagnetic \cite{electrongas}.

  Let us now consider the case of transition metals.
  In eqn.~(\ref{eqn_ferromagnetic_ratio_rule_alternative}), $n_B$ is 
the only quantity which is measured unambiguously and accurately.
  $P$ is measurable, but different methods yield different results 
\cite{spin_asymmetry_A,spin_asymmetry_B}. Furthermore, these 
experimental numbers do not match with the results of theoretical 
calculation \cite{band_calculation}.

  In principle, theoretical numbers for $g$ should be compared against 
theoretical numbers for $\rho$, and experimental numbers for $g$ should 
be compared against the experimental numbers for $\rho$.
  Let us first look at the experimental numbers.

 \begin{table}[ht]
 \begin{tabular}{ccccc}
 \hline
 element & carriers & $P_T$ & $P_C$ & $n_B$ \\
 \hline
 Fe & electrons & 0.40 & $0.42\sim0.46$ & 2.22 \\
 Co & holes & 0.35 & 0.42 & 1.72 \\
 Ni & holes & 0.23 & $0.43\sim0.465$ & 0.606 \\
 \hline
 \end{tabular}

 \vskip 10pt

 \begin{tabular}{cccc}
 \hline
 element & $n=n_B/P$ & $n_\mathrm{nominal}$ & $n_H$ \\
 \hline
 Fe & $4.8\sim5.6$ & 8 & 3.00 \\
 Co & $3.7\sim4.9$ & 3 & 0.520 \\
 Ni & $1.3\sim2.6$ & 2 & 1.12 \\
 \hline
 \end{tabular}
 \caption{\label{table_spin_asymmetry_exp} Experimental numbers for spin 
asymmetry $P$, magnetic moment $n_B$ per site, $n$ calculated as the 
ratio of these two, the nominal number of charge carriers, and the 
number of charge carriers as measured using the Hall effect.
  Two values for $P$ are taken from refs.~\cite{spin_asymmetry_A} and 
~\cite{spin_asymmetry_B}, respectively. $n_B$ is from 
ref.~\cite{Kittel}. $n_H$ is calculated from $R_H$ listed in 
ref.~\cite{Hall_effect}.
 }
 \end{table}

  In tab.~\ref{table_spin_asymmetry_exp}, we summarize the experimental 
numbers for $P$, $n_B$ and $n$. The calculated values of $n$ are 
compared against the nominal number of charge carriers, i.e., the number 
of 4s and 3d electrons or holes, and with the number of carriers 
calculated from the Hall ratio $R_H=-1/ne$. We see that the the values 
of $n$ do not seem to be in contradiction of the nominal number of 
charge carriers, in the sense that $n<n_\mathrm{nominal}$ for Fe and 
$n\approx n_\mathrm{nominal}$ for Ni and, arguably, Co.
  However, the variation in the experimental numbers is too great to 
make a concrete statement.

  Let us now turn to the theoretical numbers.

 \begin{table}
 \begin{tabular}{ccccc}
 \hline
  element & $g_\downarrow/g_\uparrow$ & $\rho_\downarrow/\rho_\uparrow$ 
  & $P$ & $n=n_B/P$ \\
 \hline
  Fe & $2.1\pm0.2$ & $2\sim2.5$ & $0.35\pm0.04$ & $6.3\pm0.7$\\
  Co & $7.0\pm1.5$ & $>1$ & $0.75\pm0.04$ & $2.3\pm0.1$ \\
  Ni & $10\pm1.5$ & $>1$ & $0.82\pm0.03$ & $0.74\pm0.03$ \\
 \hline
 \end{tabular}
 \caption{\label{table_spin_asymmetry_the} Theoretical numbers for 
$g_\downarrow/g_\uparrow$ and $\rho_\downarrow/\rho_\uparrow$. The 
numbers for $g$ were measured using a ruler applied to the 
density-of-state plots of ref.~\cite{band_calculation}. The numbers for 
$\rho$ were estimated by the eye and a ruler. The Fermi level is at the 
tail of the density of states for the case of Co and Ni, and rendering 
the definition of $\rho$ difficult or arbitrary. We also show the values 
of $P$ corresponding to these values of $g_\downarrow/g_\uparrow$, and 
an estimate of $n$ based on the actual values of $n_B$ which are shown 
in tab.~\ref{table_spin_asymmetry_exp}.}
 \end{table}

  In tab.~\ref{table_spin_asymmetry_the}, we show the numbers for 
$g_\downarrow/g_\uparrow$ as estimated from 
ref.~\cite{band_calculation}.
  In the case of Fe, $\rho$ could be estimated roughly by the eye as the 
ratio of the areas underneath the density-of-state curves. For Co and 
Ni, this was not possible because of the long tails in these curves. 
However, the relative size of $g$ was consistent with the nature of the 
carriers. That is, the density of states was found to be greater for the 
majority spin.
  All three cases are thus not inconsistent with 
eqn.~(\ref{eqn_ferromagnetic_ratio_rule_simple}).

  We also show the values of $P$ calculated from 
$g_\downarrow/g_\uparrow$. These are quite different from the 
experimental numbers introduced earlier. As a result, the numbers for 
$n$ differ from before, if we use the same values of $n_B$.

  To summarize, it is difficult to check the ratio rules quantitatively, 
at the present level of accuracy.

 \subsection{Physical interpretation}

  Let us discuss tadpole cancellation in a more intuitive fashion.

 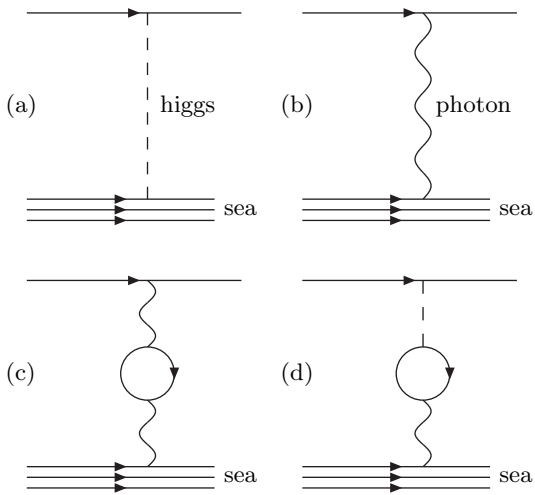
\begin{figure}[ht]
  \begin{picture}(100,100)(0,0)
   \Text(2,45)[l]{(a)}
   \ArrowLine(10,10)(80,10)
   \ArrowLine(10,6)(80,6)
   \ArrowLine(10,2)(80,2)
   \Text(90,6)[c]{sea}
   \ArrowLine(10,80)(90,80) 
   \DashLine(55,80)(55,10){5}
   \Text(60,45)[l]{higgs}
  \end{picture}
  \begin{picture}(100,100)(0,0)
   \Text(2,45)[l]{(b)}
   \ArrowLine(10,10)(80,10)
   \ArrowLine(10,6)(80,6)
   \ArrowLine(10,2)(80,2)
   \Text(90,6)[c]{sea}
   \ArrowLine(10,80)(90,80) 
   \Photon(55,80)(55,10){3}{3.5}
   \Text(60,45)[l]{photon}
  \end{picture}

  \begin{picture}(100,100)(0,0)
   \Text(2,45)[l]{(c)}
   \ArrowLine(10,10)(80,10)
   \ArrowLine(10,6)(80,6)
   \ArrowLine(10,2)(80,2)
   \Text(90,6)[c]{sea}
   \ArrowLine(10,80)(90,80) 
   \Photon(55,80)(55,55){3}{1.5}
   \Photon(55,35)(55,10){3}{1.5}
   \ArrowArcn(55,45)(10,180,-180)
  \end{picture}
  \begin{picture}(100,100)(0,0)
   \Text(2,45)[l]{(d)}
   \ArrowLine(10,10)(80,10)
   \ArrowLine(10,6)(80,6)
   \ArrowLine(10,2)(80,2)
   \Text(90,6)[c]{sea}
   \ArrowLine(10,80)(90,80) 
   \DashLine(55,80)(55,55){5}
   \Photon(55,35)(55,10){3}{1.5}
   \ArrowArcn(55,45)(10,180,-180)
  \end{picture}
 \caption{\label{fig_ratio_rule}
  The interaction of a conduction electron with the Fermi sea of 
electrons.
 }
 \end{figure}

  Figure \ref{fig_ratio_rule} represents the interaction experienced by 
a conduction electron or holes due to the surrounding Fermi sea of 
(conduction) electrons or holes. Let us say that the charge carriers are 
electrons. Note that fig.~\ref{fig_ratio_rule}a corresponds to 
eqn.~(\ref{eqn_higgs_tadpole}), and fig.~\ref{fig_ratio_rule}d 
corresponds to eqn.~(\ref{eqn_photon_tadpole}).

  The interaction shown in Fig.~\ref{fig_ratio_rule}a involves the 
exchange of the density fluctuation of spin, i.e., the Higgs boson. This 
boson being a scalar, its exchange is always repulsive between like 
particles, i.e., between the same spin.
  Since, by definition, there are more majority-spin electrons than 
minority spin electrons, this interaction makes majority-spin electrons 
more energetically unfavourable. That is, Fig.~\ref{fig_ratio_rule}a, or 
the density fluctuation of spin, tends to suppress magnetic order.

  The exchange of the Higgs boson is not the only interaction between 
the conduction electrons and the sea, and in Fig.~\ref{fig_ratio_rule}b, 
we show the Coulomb exchange. This is always repulsive, and is of the 
same magnitude for both type of electrons, and so this diagram does not 
contribute to the formation or suppression of magnetic order.

  Figure \ref{fig_ratio_rule}b by itself is infinite since the photon 
propagator diverges at zero momentum transfer. This is, as usual, 
remedied by the screening effect which is shown in 
fig.~\ref{fig_ratio_rule}c.

  The screening effect, such as that shown in 
fig.~\ref{fig_ratio_rule}c, usually suppresses the charge. This is 
because a negative charge attracts positive charge, and this positive 
charge tends to cancel the negative charge.

  However, the contribution of fig.~\ref{fig_ratio_rule}d requires more 
thought.
  The sea electrons, which have negative charge, attracts positive 
charge.
  When this positive charge has the same spin as the conduction 
electron, i.e., when the positive charge suppresses the electronic spin 
which is aligned with the spin of the conduction electron, the 
positive charge attracts this conduction electron.
  On the other hand, when the positive charge has opposite spin to that 
of the conduction electron, then the conduction electron is repelled.

  Whether the interaction of fig.~\ref{fig_ratio_rule}d tends to create 
magnetic order or suppress it depends on which type of spin is more 
likely to be excited, i.e., on the density of states at the Fermi 
surface. This is the meaning of the tadpole cancellation.
  In other words, the ferromagnetic ground state is stable when the 
interaction due to the fluctuation of spin density, which is mediated by 
the Higgs boson and which always suppresses the polarization of spin, is 
equal and opposite to the contribution due to the electrostatic 
polarization of fig.~\ref{fig_ratio_rule}d which, depending on 
circumstances, can counteract it.

 \subsection{Comparison with the Hubbard model}

  When the Coulomb interaction is screened, the interaction becomes 
point-like in the limit of large screening, i.e.,
 \begin{equation}
  \frac1{g_\downarrow(\epsilon_F)+g_\uparrow(\epsilon_F)}
  \longrightarrow \hat{U},
 \end{equation}
  where $\hat U$ is a constant which can be interpreted as the on-site 
Coulomb repulsion $U$ up to a normalization.
  Let us now see what would happen if we were to start from a theory 
which treats the on-site Coulomb repulsion $U$ as the starting point, 
such as the Hubbard model.

  In this case, eqn.~(\ref{eqn_higgs_tadpole}) is unchanged, but 
eqn.~(\ref{eqn_photon_tadpole}) is modified to take the following form:
 \begin{equation}
  \begin{picture}(115,30)(0,0)
   \ArrowArcn(50,-36)(60,120,60)
   \ArrowArcn(50,66)(60,300,240)
   \GCirc(20,15){3}{0}
   \Text(8,15)[c]{$\sigma^3_{ab}$}
   \Text(85,15)[l]{$\hat U$}
   \GCirc(80,15){3}{0}
   \ArrowArcn(95,15)(15,270,-90)
  \end{picture}.
  \label{eqn_photon_tadpole_hubbard}
 \end{equation}
  Here, as is usual in the Hubbard model, the spin going into the fish 
part must be opposite to the spin going into the tadpole. That is,
 \begin{equation}
  \mathcal{A}^\mathrm{tadpole}_\mathrm{Hubbard}=
  \hat U\left[g_\uparrow(\epsilon_F)\rho_\downarrow-
  g_\downarrow(\epsilon_F)\rho_\uparrow\right],
  \label{eqn_photon_tadpole_hubbard_amplitude}
 \end{equation}
  in the place of eqn.~(\ref{eqn_photon_tadpole_amplitude}).
  This would lead to different consequences.

  The origin of this discrepancy is clear. Expressed in terms of the 
screened Coulomb propagator, there are two contributions that go into 
eqn.~(\ref{eqn_photon_tadpole_hubbard}). One is the genuine tadpole-like 
contribution of the form eqn.~(\ref{eqn_photon_tadpole}).
  The contribution of this term is given by
 \begin{equation}
  \hat U\left[g_\uparrow(\epsilon_F)-g_\downarrow(\epsilon_F)\right]
  \left(\rho_\downarrow+\rho_\uparrow\right),
 \end{equation}
  so that this term has the same form as in 
eqn.~(\ref{eqn_photon_tadpole_hubbard}).
  On the other hand, there is a second contribution, which is the 
self-energy correction:
 \begin{equation}
  \begin{picture}(55,30)(0,0)
   \ArrowArcn(35,15)(15,180,-180)
   \Photon(35,0)(35,30){3}{2}
   \GCirc(20,15){3}{0}
   \Text(8,15)[c]{$\sigma^3_{ab}$}
  \end{picture}.
 \end{equation}
  The contribution due to this term is given by
 \begin{equation}
  \hat U\left[g_\downarrow(\epsilon_F)\rho_\downarrow-
  g_\uparrow(\epsilon_F)\rho_\uparrow\right].
 \end{equation}
  Adding together these two contributions yields 
eqn.~(\ref{eqn_photon_tadpole_hubbard_amplitude}).

  The discrepancy comes because in our approach, the self-energy 
correction is absorbed in the all-order Green's function, whereas in the 
Hubbard-model approach, this is not possible. In the Hubbard model, 
either both contributions are treated as a tadpole, or both 
contributions are treated as a self-energy correction.
  If the latter, one will have the condition that the simple tadpole, 
with the self-energy corrections, by itself vanishes. This condition 
requires $\rho_\uparrow=\rho_\downarrow$, and therefore we will never 
have a stable ferromagnetic solution out of the Hubbard model.

  This, in our opinion, is a limitation of the Hubbard model.
  The limitation is due to the inability to treat current--current 
mixing, which is the basis of our discussion in this section.

  One may still argue that on-site Coulomb repulsion is present, 
physically.
  In other words, the effective screened Coulomb propagator, which 
ordinarily gives a divergent contribution at the origin up to a UV 
cut-off, is not really divergent but only large and finite at the 
origin.

  If so, this may be thought of as a variant of the UV cut-off of the 
Coulomb propagator $D_\mathrm{photon}$, which may be parametrized, for 
example, as
 \begin{equation}
  D_\mathrm{photon}(\mathbf{k})=\frac1{-\mathbf{k}^2-a\mathbf{k}^4},
 \end{equation}
  where $a$ is a parameter (positive or negative).
  Even if this is not permissible as a field theory, it is permissible 
as an UV (Pauli--Villars) regularization procedure.
  It will be seen that such a cut-off does not affect our argument at 
all, since our discussion involves zero momentum photons.
  The electron self-energy will be affected by the UV cut-off, but this 
does not affect our results explicitly.

 \section{Analysis of spin current conservation}
 \label{sec_spin_current}

  Let us now move on to the formalism, which is required if we are to go 
beyond the tadpole-level analysis of the preceding section.
  We adapt Gribov's analysis of axial current conservation 
\cite{gribov,gribovewsb} to the context of spin current conservation in 
systems with partial magnetic order.

  To sum up in one phrase, our goal is to start from the Coulombic 
system, which is defined by eqn.~(\ref{eqn_Coulombic_Lagrangian}), and 
solve it as exactly as possible using the Dyson--Schwinger equations, 
under a number of assumptions.

  The major assumption is that of spontaneous symmetry breaking. If the 
spin symmetry is broken spontaneously, then the Goldstone theorem 
guarantees the presence of Goldstone and Higgs modes. These modes are, 
in terms of the initial Lagrangian, electronic excitations.
  However, in terms of the effective theory that appears at the end, 
they are elementary, and participate in the conservation of the spin 
current.
  This is the main property that allows us to solve the Dyson--Schwinger 
equations. The other assumptions, such as the linear or quadratic form 
of the magnon dispersion relation and the constancy of exchange energy 
which are sometimes required, are approximations, which we believe are 
viable, that can be lifted if one has the computational resources.

  The resulting effective Lagrangian is found to have the following 
interaction term:
 \begin{equation}
  \mathcal{L}^\mathrm{I}_\mathrm{eff}
  \propto\psi^\dagger\Phi\cdot\sigma\psi.
  \label{eqn_effective_lagrangian}
 \end{equation}
  Here $\Phi^i$ is essentially the order-parameter field, but with a 
certain formal difference which we shall discuss.
  We would like to emphasize at this point that this equation is not our 
starting point.
  It is rather the end product of solving the Coulombic system by means 
of the Dyson--Schwinger equations, with the aid of the Goldstone theorem 
and the Ward--Takahashi identities.

  Let us start by discussing current conservation. As discussed in the 
previous section, current is absolutely conserved when the spin symmetry 
is conserved.
 \begin{equation}
  \frac{\partial}{\partial x^\mu}J^{\mu,i}_\mathrm{spin}=0.
 \end{equation}
  
  Current conservation is reflected in the following Ward--Takahashi 
identity:
 \begin{equation}
  \Gamma^\mu (q_1-q_2)_\mu=
  G^{-1}_{\lambda_1}(q_1)-G^{-1}_{\lambda_2}(q_2).
  \label{eqn_ward_takahashi}
 \end{equation}
  $\Gamma^\mu$ is the vertex in the momentum space.
  $\lambda_{1,2}$ refer to the spin states, but these are dummy indices 
here in the sense that $G^{-1}_\lambda(q)$ is independent of $\lambda$. 
Thus eqn.~(\ref{eqn_ward_takahashi}) holds for any combination of spin, 
and therefore current is conserved.
  $q$ are $d+1$-vectors with components $(q_0,\mathbf{q})$.
  $q_0$ is the energy and $\mathbf{q}$ is the spatial momentum, with 
$\hbar=1$.

  The Ward--Takahashi identity is violated in the symmetry-broken phase, 
since there is now an energy difference $\Delta E$, which is the 
exchange energy, between the different spin states:
 \begin{equation}
  \Delta E= G^{-1}_\downarrow(q)-G^{-1}_\uparrow(q).
  \label{eqn_delta_E_definition}
 \end{equation}
  This is the definition for the ferromagnetic case, when $\Delta E$ is 
positive if we take $\downarrow$ to be the majority spin state.
  In the anti-ferromagnetic case, $\Delta E$ is positive in one 
sub-lattice and negative in the other.

  If $G^{-1}_\downarrow$ and $G^{-1}_\uparrow$ are both linear in 
energy, $\Delta E$ is given by $\epsilon_\uparrow-\epsilon_\downarrow$ 
and is constant up to a possible dependence on the spatial momentum 
$\mathbf{q}$.
  In principle, $\Delta E$ depends on $q_0$ and $\mathbf{q}$.
  In particular, at the threshold, $G^{-1}$ would, in general, have a 
singular structure corresponding to the emission and absorption of the 
Goldstone boson $\phi$ through the process $e_*\to e\phi$.
  The results of the previous section are stable against such 
corrections, as we have discussed. However, the results of this section 
are more easily derived for constant $\Delta E$, which corresponds to 
the case of the Fermi liquid whose exchange energy is constant.

  After the symmetry violation, the currents $J^\mu_{1,2}$ are no longer 
conserved, and the Ward--Takahashi identity is violated by
 \begin{equation}
  \Gamma^\mu (q_1-q_2)_\mu\propto
  G^{-1}_{\lambda_1}(q_1)-G^{-1}_{\lambda_2}(q_2)
  \pm\Delta E\quad (\lambda_1\ne\lambda_2).
  \label{eqn_ward_takahashi_violation}
 \end{equation}
  This $\Delta E$ contribution is of the same form as the coupling of 
the Goldstone boson $\phi$, and current conservation is restored by 
including the contribution of the Goldstone boson. This is the case even 
when $\Delta E$ is not constant.
  Specifically, spin current conservation is restored for the vertex 
$\widetilde\Gamma$ which is modified by the inclusion of the Goldstone 
boson,
 \begin{equation}
  \begin{picture}(155,50)(0,0)
   \Text(5,27)[c]{$\widetilde{\Gamma}^\mu=$}
   \ArrowLine(25,25)(50,25)
   \Line(47,20)(53,30) \Line(53,20)(47,30)
   \Text(50,32)[b]{$\mu$}
   \ArrowLine(50,25)(75,25)
   \Text(85,27)[c]{$+$}
   \ArrowLine(95,5)(120,15)
   \ArrowLine(120,15)(145,5)
   \DashLine(120,15)(120,35){5}
   \Line(117,30)(123,40) \Line(123,30)(117,40)
   \Text(120,42)[b]{$\mu$}
  \end{picture}.
  \label{eqn_modified_vertex}
 \end{equation}
  As before, the crosses indicate the spin-current vertices, and the 
dashed line indicates the Goldstone boson.
  There is nothing strange in this result, since the Goldstone boson 
arose in the first place as the longitudinal component of the spin 
current.
  After taking away the longitudinal component, the remaining part is 
transverse and therefore satisfies the Ward--Takahashi identity.
  The current--magnon two-point function which appears in the second 
term consists of fermionic and bosonic loop. The latter contains magnons 
and the Higgs boson.

  The Goldstone bosons $\phi_1$ and $\phi_2$ correspond to the SU(2) 
rotation perpendicular to the local orientation of spin (which is along 
$z$),
 \begin{equation}
  \mathcal{U}(\phi_1,\phi_2)=\exp
  \left[if^{-1}\sum_{i=1}^{2}\phi_i\sigma_i\right],
  \label{eqn_goldstone_rotation}
 \end{equation}
  and they correspond physically to the magnons.
  $f$ is the Goldstone boson form factor which, by virtue of 
eqn.~(\ref{eqn_modified_vertex}), is calculated as the strength of the 
current--Goldstone-boson two-point amplitude.


 \subsection{The two-point function and the coupling with fermions}

  We have noted in eqn.~(\ref{eqn_SU_two_breaking}) that there is a 
residual symmetry associated with U(1)$_z$, which is conserved.
  The states can be classified according to the charges under this 
rotation group. First, we define the charge of $\psi_\uparrow$ to be 
$+1/2$. The remaining charges follow automatically, and we obtain the 
values listed in tab.~\ref{tab_fields_and_charges}. These are 
necessarily conserved.
  Note that the U(1)$_\mathrm{EM}$ charges are $e$ for the electron/hole 
fields and 0 for all others. These charges are also conserved.

 \begin{table}[ht]
  \begin{tabular}{ccc}
   \hline
   field & U(1)$_z$ charge & U(1)$_\mathrm{EM}$ charge \\
   \hline
   $\psi_+\equiv\psi_\uparrow$ & $+1/2$ & $e$ \\
   $\psi_-\equiv\psi_\downarrow$ & $-1/2$ & $e$ \\
   $\phi_+\equiv-\phi_1+i\phi_2$ & $+1$ & 0 \\
   $\phi_-\equiv \phi_1+i\phi_2$ & $-1$ & 0 \\
   $h_0\equiv h$ & $0$ & 0 \\
   \hline
  \end{tabular}
  \caption{\label{tab_fields_and_charges} The U(1)$_z$ and 
U(1)$_\mathrm{EM}$ charges of the fields. $e$ is positive for holes and 
negative for electrons.}
 \end{table}

  In order that the Ward--Takahashi identity is satisfied by the vertex 
of eqn.~(\ref{eqn_modified_vertex}), the following identity needs to be 
satisfied:
 \begin{equation}
  \begin{picture}(150,40)(0,0)
   \Line(27,15)(33,25) \Line(33,15)(27,25)
   \DashLine(30,20)(50,20){5}
   \Text(25,20)[r]{$\mu$}
   \Text(55,20)[l]{$\times q^\mu=-fD^{-1}_\phi(q)$.}
  \end{picture}
  \label{eqn_two_point_Ward_identity}
 \end{equation}
  Here, $f$ is a constant of proportionality, and is the same quantity 
as that which appears in eqn.~(\ref{eqn_goldstone_rotation}).
  $q^\mu$ is the momentum flowing into the two-point function from the 
current (i.e., left to right).
  This present definition of $f$ is more rigorous.
  Note that this also fixes the sign of $D_\phi$. Our present definition 
corresponds to taking the couplings to be real and taking the sign of 
$D_\phi$ to be opposite to that for scalar particles.

\begin{figure}[ht]
  \begin{picture}(100,80)(0,0)
   \ArrowLine(25,20)(50,35)
   \ArrowLine(50,35)(75,20)
   \DashLine(50,35)(50,60){5}
   \LongArrow(55,60)(55,50)
   \Text(27,65)[c]{(a)}
   \Text(25,27)[b]{$+1/2$}
   \Text(75,27)[b]{$-1/2$}
   \Text(55,65)[b]{$-1$}
   \Text(50,5)[b]{$+f^{-1}\Delta E$}
  \end{picture}
  \begin{picture}(100,80)(0,0)
   \ArrowLine(25,20)(50,35)
   \ArrowLine(50,35)(75,20)
   \DashLine(50,35)(50,60){5}
   \LongArrow(55,60)(55,50)
   \Text(27,65)[c]{(b)}
   \Text(25,27)[b]{$-1/2$}
   \Text(75,27)[b]{$+1/2$}
   \Text(55,65)[b]{$+1$}
   \Text(50,5)[b]{$-f^{-1}\Delta E$}
  \end{picture}

  \begin{picture}(100,80)(0,0)
   \ArrowLine(25,20)(50,35)
   \ArrowLine(50,35)(75,20)
   \DashLine(50,35)(50,60){5}
   \LongArrow(55,60)(55,50)
   \Text(27,65)[c]{(c)}
   \Text(25,27)[b]{$+1/2$}
   \Text(75,27)[b]{$+1/2$}
   \Text(55,65)[b]{$0$}
   \Text(50,5)[b]{$+f^{-1}\Delta E$}
  \end{picture}
  \begin{picture}(100,80)(0,0)
   \ArrowLine(25,20)(50,35)
   \ArrowLine(50,35)(75,20)
   \DashLine(50,35)(50,60){5}
   \LongArrow(55,60)(55,50)
   \Text(27,65)[c]{(c)}
   \Text(25,27)[b]{$-1/2$}
   \Text(75,27)[b]{$-1/2$}
   \Text(55,65)[b]{$0$}
   \Text(50,5)[b]{$-f^{-1}\Delta E$}
  \end{picture}
  \caption{\label{fig_boson_fermion_vertices}
  The Feynman rules for the coupling of the magnons and the Higgs boson 
with the fermions.}
 \end{figure}
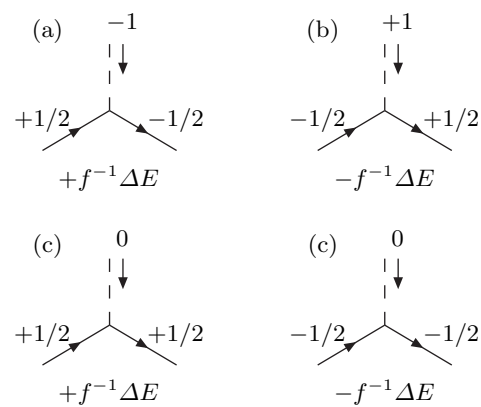

  Given this definition of $f$, we can determine the coupling constants 
with the fermions by the condition that eqn.~(\ref{eqn_modified_vertex}) 
satisfies the Ward--Takahashi identity. We then obtain the Feynman rules 
that are given in figs.~\ref{fig_boson_fermion_vertices}a and b.

  For example, for the configuration of 
fig.~\ref{fig_boson_fermion_vertices}a, eqn.~(\ref{eqn_modified_vertex}) 
yields the following Ward--Takahashi identity:
 \begin{equation}
  q^\mu\Gamma_\mu+i^2(-fD_\phi^{-1}(q))D_\phi(q)(f^{-1}\Delta E)
  =G_-^{-1}-G_+^{-1}.
 \end{equation}
  Note that the same Feynman rules can be obtained, less rigorously, by 
considering the rotation associated with the Goldstone bosons, 
eqn.~(\ref{eqn_goldstone_rotation}), and considering its coupling with 
the fermions in eqn.~(\ref{eqn_Coulombic_Lagrangian}).

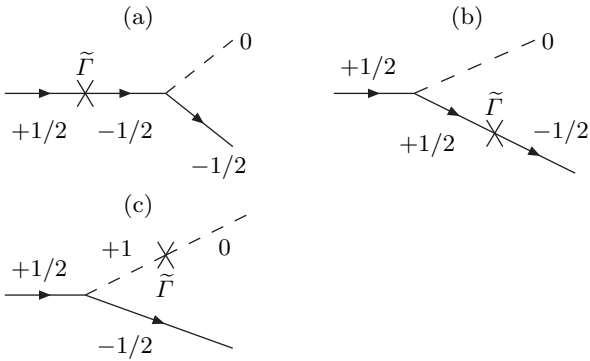
\begin{figure}[ht]
  \begin{picture}(120,70)(0,0)
   \Text(60,68)[t]{(a)}
   \ArrowLine(10,35)(40,35)
   \Line(37,30)(43,40) \Line(37,40)(43,30)
   \ArrowLine(40,35)(70,35)
   \ArrowLine(70,35)(95,15)
   \DashLine(70,35)(95,55){5}
   \Text(40,42)[b]{$\widetilde\Gamma$}
   \Text(23,25)[t]{$+1/2$}
   \Text(55,25)[t]{$-1/2$}
   \Text(90,12)[t]{$-1/2$}
   \Text(98,55)[l]{$0$}
  \end{picture}
  \begin{picture}(120,70)(0,0)
   \Text(60,68)[t]{(b)}
   \ArrowLine(10,35)(40,35)
   \DashLine(40,35)(85,55){5}
   \ArrowLine(40,35)(70,20)
   \Line(67,15)(73,25) \Line(67,25)(73,15)
   \ArrowLine(70,20)(100,5)
   \Text(70,27)[b]{$\widetilde\Gamma$}
   \Text(23,40)[b]{$+1/2$}
   \Text(45,20)[t]{$+1/2$}
   \Text(95,25)[t]{$-1/2$}
   \Text(88,55)[l]{$0$}
  \end{picture}
 
 \begin{picture}(120,70)(0,0)
   \Text(60,68)[t]{(c)}
   \ArrowLine(10,30)(40,30)
   \DashLine(40,30)(70,45){5}
   \Line(67,40)(73,50) \Line(67,50)(73,40)
   \DashLine(70,45)(100,60){5}
   \ArrowLine(40,30)(95,10)
   \Text(70,38)[t]{$\widetilde\Gamma$}
   \Text(23,35)[b]{$+1/2$}
   \Text(55,15)[t]{$-1/2$}
   \Text(52,44)[b]{$+1$}
   \Text(90,48)[l]{$0$}
  \end{picture}
  \caption{\label{fig_fermion_scalar_current_amplitude}
  The three diagrams whose sum must satisfy the Ward--Takahashi 
identity. The crosses correspond to the modified current vertex defined 
by eqn.~(\ref{eqn_modified_vertex}).}
 \end{figure}

  The vertices that involve the Higgs boson, which are shown in 
fig.~\ref{fig_boson_fermion_vertices}c and d, cannot be fixed by this 
particular type of Ward--Takahashi identity. However, they can be fixed 
by considering the current insertion in the three-point amplitude, for 
example, as shown in fig.~\ref{fig_fermion_scalar_current_amplitude}.

  The Ward--Takahashi identity applied to 
fig.~\ref{fig_fermion_scalar_current_amplitude} also allows us to 
determine the magnon--magnon--Higgs vertex.
  However, for doing so, we need to know the form of $D_\phi(q)$ and 
$D_h(q)$.
  Let us therefore calculate the current--magnon two-point function of 
eqn.~(\ref{eqn_two_point_Ward_identity}).

 \begin{figure}[ht]
  \begin{picture}(130,50)(0,0)
   \Text(20,48)[t]{(a)}
   \ArrowArcn(50,-36)(60,120,60)
   \ArrowArcn(50,66)(60,300,240)
   \Line(17,10)(23,20) \Line(23,10)(17,20)
   \GCirc(80,15){3}{0}
   \Text(20,21)[b]{$\mu$}
   \LongArrow(5,15)(15,15)
   \Text(10,14)[t]{$q$}
   \Text(90,15)[l]{$f^{-1}\Delta E$}
   \Text(50,25)[b]{$+1/2$}
   \Text(50,5)[t]{$-1/2$}
  \end{picture}
  \begin{picture}(100,30)(0,0)
   \Text(20,48)[t]{(b)}
   \DashCArc(50,-36)(60,60,120){5}
   \DashCArc(50,66)(60,240,300){5}
   \Line(17,10)(23,20) \Line(23,10)(17,20)
   \GCirc(80,15){3}{0}
   \Text(20,21)[b]{$\mu$}
   \LongArrow(5,15)(15,15)
   \Text(10,14)[t]{$q$}
   \Text(50,25)[b]{$+1$}
   \Text(50,5)[t]{$0$}
  \end{picture}
  \caption{\label{fig_two_point_func_loop}
  The fermionic (a) and bosonic (b) contributions to the current--magnon 
two-point function.}
 \end{figure}
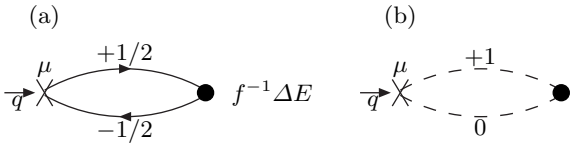

  For now, we calculate the fermionic loop, which is shown in 
fig.~\ref{fig_two_point_func_loop}a.
  It should be noted that the end result of this calculation is 
independent of whether we consider $\phi_+$ or $\phi_-$.
  Using the Feynman rule of fig.~\ref{fig_boson_fermion_vertices}a, we 
obtain
 \begin{eqnarray}
  &&\hspace{-10pt}i\mathcal{A}^\mu_\mathrm{two-point}(q)=\nonumber\\
  &&i^4(-1)\int\frac{d^{d+1}k}{(2\pi)^{d+1}}
  \Gamma^\mu G_+(k)G_-(k-q)(f^{-1}\Delta E).
  \label{eqn_zero_momentum_two_point_amplitude}
 \end{eqnarray}
  In particular, for the case $q\to0$, we obtain the exact expression:
 \begin{equation}
  \mathcal{A}^\mu_\mathrm{two-point}(q\to0)=
  \int\frac{d^{d+1}k}{(2\pi)^{d+1}i}
  f^{-1}\Gamma^\mu \left[G_-(k)-G_+(k)\right],
 \end{equation}
  where we made use of eqn.~(\ref{eqn_delta_E_definition}). For Fermi 
liquids, $\Gamma^0=1$ and, by symmetry, the spatial components of this 
amplitude usually vanishes at $q=0$. We thus obtain
 \begin{equation}
  \mathcal{A}^\mu_\mathrm{two-point}(q\to0)=
  f^{-1}(\rho_--\rho_+,0,0,0).
  \label{eqn_zero_momentum_two_point_amplitude_result}
 \end{equation}
  Note that this vanishes for the case of anti-ferromagnetism where 
there is no global spin asymmetry.

  When we compare 
eqn.~(\ref{eqn_zero_momentum_two_point_amplitude_result}) with 
eqn.~(\ref{eqn_two_point_Ward_identity}), we see immediately that
 \begin{equation}
  f^2=\rho_--\rho_+ \quad \mbox{(ferromagnetism),}
 \end{equation}
  and
 \begin{equation}
  \left\{
   \begin{array}{ll}
    -D_\phi^{-1}(q)=q_0-\mathbf{q}^2/2m_\phi &
    \mbox{(ferromagnetism),}\\
    -D_\phi^{-1}(q)=q_0^2-u^2\mathbf{q}^2 & 
    \mbox{(anti-ferromagnetism),}
   \end{array}
  \right.
 \end{equation}
  for small energy and momenta. The inclusion of the bosonic loop does 
not alter this conclusion.
  The unusual negative sign of $D$ reflects the fact that the magnons 
are pseudo-scalar. That is, the fields are $i\phi$ rather than $\phi$ in 
our convention.

  We need to calculate $f^2$ by other means, such as calculating the 
two-point amplitude for finite $q$, for anti-ferromagnetism. $u$ and 
$m_\phi$ are parameters which are in principle calculable by, for 
example, evaluating the finite $q$ case.
  However, the form of eqn.~(\ref{eqn_zero_momentum_two_point_amplitude}) 
implies $m_\phi\sim m_e$ and $u\sim v_F$.

  There is actually a smarter method than to calculate the finite-$q$ 
case (which is cumbersome), but the full calculation, in the case of 
anti-ferromagnetism, requires our knowledge of the bosonic three-point 
functions. Let us therefore postpone the calculation of 
anti-ferromagnetic $f^2$ and other parameters for now.

 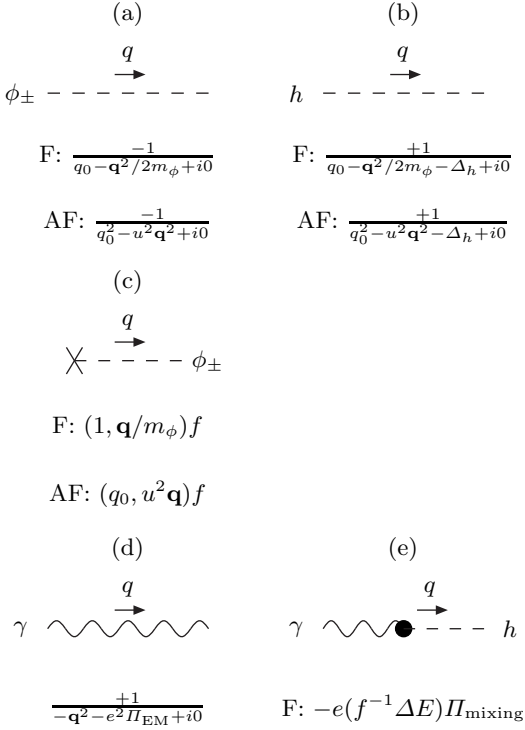
\begin{figure}[ht]
  \begin{picture}(100,100)(0,0)
   \Text(50,90)[c]{(a)}
   \DashLine(20,60)(80,60){5}
   \LongArrow(45,67)(55,67)
   \Text(50,72)[b]{$q$}
   \Text(10,60)[c]{$\phi_\pm$}
   \Text(50,35)[c]{F: $\frac{-1}{q_0-\mathbf{q}^2/2m_\phi+i0}$}
   \Text(50,10)[c]{AF: $\frac{-1}{q_0^2-u^2\mathbf{q}^2+i0}$}
  \end{picture}
  \begin{picture}(100,100)(0,0)
   \Text(50,90)[c]{(b)}
   \DashLine(20,60)(80,60){5}
   \LongArrow(45,67)(55,67)
   \Text(50,72)[b]{$q$}
   \Text(10,60)[c]{$h$}
   \Text(50,35)[c]{F: $\frac{+1}{q_0-\mathbf{q}^2/2m_\phi-\Delta_h+i0}$}
   \Text(50,10)[c]{AF: $\frac{+1}{q_0^2-u^2\mathbf{q}^2-\Delta_h+i0}$}
  \end{picture}

  \begin{picture}(100,100)(0,0)
   \Text(50,90)[c]{(c)}
   \Line(27,55)(33,65) \Line(33,55)(27,65)
   \DashLine(30,60)(70,60){5}
   \LongArrow(45,67)(55,67)
   \Text(50,72)[b]{$q$}
   \Text(80,60)[c]{$\phi_\pm$}
   \Text(50,35)[c]{F: $(1,\mathbf{q}/m_\phi)f$}
   \Text(50,10)[c]{AF: $(q_0,u^2\mathbf{q})f$}
  \end{picture}

  \begin{picture}(100,80)(0,0)
   \Text(50,70)[c]{(d)}
   \Photon(20,40)(80,40){3}{4.5}
   \LongArrow(45,47)(55,47)
   \Text(50,52)[b]{$q$}
   \Text(10,40)[c]{$\gamma$}
   \Text(50,10)[c]{$\frac{+1}{-\mathbf{q}^2-e^2\Pi_\mathrm{EM}+i0}$}
  \end{picture}
  \begin{picture}(100,80)(0,0)
   \Text(50,70)[c]{(e)}
   \Photon(20,40)(50,40){3}{2.5}
   \GCirc(50,40){3}{0}
   \DashLine(50,40)(80,40){5}
   \LongArrow(55,47)(65,47)
   \Text(60,52)[b]{$q$}
   \Text(10,40)[c]{$\gamma$}
   \Text(90,40)[c]{$h$}
   \Text(50,10)[c]{F: $-e(f^{-1}\Delta E)\Pi_\mathrm{mixing}$}
  \end{picture}
  \caption{\label{fig_boson_two_point_functions}
  The propagators and the two-point functions. $\Pi_\mathrm{EM}$ is 
given by the $0-0$ component of eqn.~(\ref{eqn_photon_self_energy}), and 
$\Pi_\mathrm{mixing}$ is given by the $0-0$ component of 
eqn.~(\ref{eqn_current_mixing_after_SSB}).}
 \end{figure}

  Let us summarize the results of this section up to here.

  Firstly, we summarize the propagators and the two-point functions in 
fig.~\ref{fig_boson_two_point_functions}.
  The Higgs-boson Green's functions are given in 
fig.~\ref{fig_boson_two_point_functions}b, with a constant energy gap 
$\Delta_h$. $\Delta_h$ is defined as the energy gap for ferromagnetism 
and the energy gap squared for anti-ferromagnetism. This definition is 
convenient when we discuss the bosonic three-point functions.
  Note that it is an approximation to say that $\Delta_h$ is independent 
of momenta and energy. However, it becomes easier to implement current 
conservation in this manner.
  For completeness's sake, we also list the screened photon Green's 
function (with the approximation that the screening, $\Pi_\mathrm{EM}$, 
is constant) and the Higgs--photon mixing. These are as given in the 
previous section.

  Secondly, the fermionic vertices are as given before in 
fig.~\ref{fig_boson_fermion_vertices}. We did not list the photonic 
vertex, but this is given by $e$.
  Note that the couplings given in fig.~\ref{fig_boson_fermion_vertices} 
can be summarized in the following compact form (c.f.\ 
eqn.~(\ref{eqn_effective_lagrangian})):
 \begin{equation}
  \mathcal{L}^\mathrm{I}_\mathrm{eff}=
  (f^{-1}\Delta E)\psi^\dagger\Phi\cdot\sigma\psi,
  \label{eqn_effective_lagrangian_with_coupling}
 \end{equation}
  where $\Phi$ is defined by
 \begin{equation}
  \Phi=(\phi_1,\phi_2,-v+h).
 \end{equation}
  $\phi_1$ and $\phi_2$ are as shown in 
tab.~\ref{tab_fields_and_charges}, and are given by
 \begin{equation}
  \phi_1=\frac12(-\phi_++\phi_-),\qquad
  \phi_2=\frac i2(\phi_++\phi_-).
 \end{equation}
  $v$ is a parameter, which has the interpretation as the vacuum 
expectation value of the $\Phi$ field.
  In order that the energy difference between the two states that is 
given by eqn.~(\ref{eqn_effective_lagrangian_with_coupling}) should agree 
with the actual energy difference $\Delta E$, $v$ needs to satisfy
 \begin{equation}
  v=f/2.
 \end{equation}
  $\Phi$ is essentially a magnetic order-parameter field. This differs 
from the more conventional form such as
 \begin{equation}
  \mathcal{U}(\phi_1,\phi_2)(0,0,v+h)^\mathrm{T},
 \end{equation}
  but they match in the limit of small fields, up to some differences in 
convention.

 \subsection{Bosonic vertices}

  Let us consider the Ward--Takahashi identity corresponding to the 
amplitude described by fig.~\ref{fig_fermion_scalar_current_amplitude}.

  We denote the initial state momentum to be $q_1$ and the final state 
momenta to be $q_2$ and $q_3$. $q_2$ is for the $h_0$ boson and $q_3$ is 
for the $-1/2$ fermion. We denote the momentum which goes into the 
vertex by $q$, so that $q+q_1=q_2+q_3$. It is not necessary that the 
fermions and the bosons are on shell, i.e., $G^{-1}_+(q_1)$ etc.\ need 
not be zero.

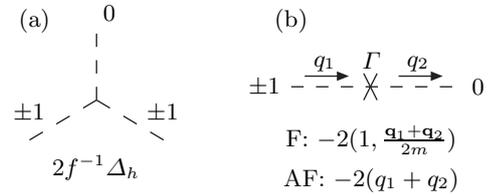
\begin{figure}[ht]
  \begin{picture}(100,80)(0,0)
   \Text(27,65)[c]{(a)}
   \DashLine(25,20)(50,35){5}
   \DashLine(50,35)(75,20){5}
   \DashLine(50,35)(50,60){5}
   \Text(25,27)[b]{$\pm1$}
   \Text(75,27)[b]{$\pm1$}
   \Text(55,65)[b]{$0$}
   \Text(50,5)[b]{$2f^{-1}\Delta_h$}
  \end{picture}
  \begin{picture}(100,80)(0,0)
   \Text(20,65)[c]{(b)}
   \DashLine(20,40)(80,40){5}
   \Line(47,35)(52,45) \Line(52,35)(47,45)
   \Text(10,40)[c]{$\pm1$}
   \Text(90,40)[c]{$0$}
   \LongArrow(25,44)(40,44)
   \LongArrow(60,44)(75,44)
   \Text(32.5,46)[b]{$q_1$}
   \Text(67.5,46)[b]{$q_2$}
   \Text(50,47)[b]{$\Gamma$}
   \Text(50,20)[c]{F: $-2(1,\frac{\mathbf{q}_1+\mathbf{q}_2}{2m})$}
   \Text(50,5)[c]{AF: $-2(q_1+q_2)$}
  \end{picture}
  \caption{\label{fig_boson_three_point_functions}
  The bosonic three-point functions. In (b), $q_1+q_2$ is a short-hand 
notation for $((q_1+q_2)_0,u^2(\mathbf{q}_1+\mathbf{q}_2))$.}
 \end{figure}

  Upon contraction with $q$, the first two diagrams yield
 \begin{equation}
  q_\mu \mathcal{A}^\mu_\mathrm{a}=
  -i^2f^{-1}\Delta E\left(1-G^{-1}_+(q_1)G_-(q+q_1)\right)
 \end{equation}
  and
 \begin{equation}
  q_\mu \mathcal{A}^\mu_\mathrm{b}=
  i^2f^{-1}\Delta E\left(G^{-1}_-(q_3)G_+(q_3-q)-1\right).
 \end{equation}
  The amplitude as a whole satisfies the Ward identity if the third 
amplitude satisfies
 \begin{equation}
  q_\mu \mathcal{A}^\mu_\mathrm{c}=
  2i^2f^{-1}\Delta E\left(1+D^{-1}_h(q_2)D_\phi(q_2-q)\right).
 \end{equation}
  This requires vertices of the form shown in 
fig.~\ref{fig_boson_three_point_functions}.

 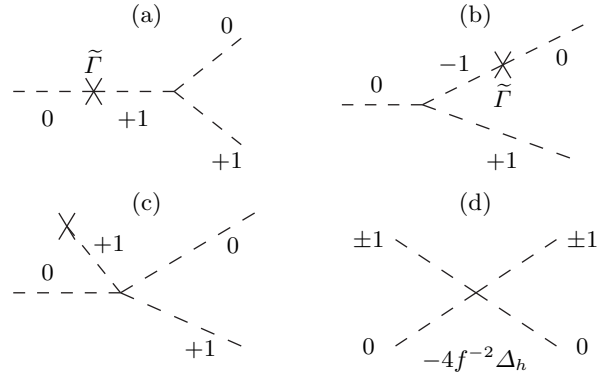
\begin{figure}[ht]
  \begin{picture}(120,70)(0,0)
   \Text(60,68)[t]{(a)}
   \DashLine(10,35)(40,35){5}
   \Line(37,30)(43,40) \Line(37,40)(43,30)
   \DashLine(40,35)(70,35){5}
   \DashLine(70,35)(95,15){5}
   \DashLine(70,35)(95,55){5}
   \Text(40,42)[b]{$\widetilde\Gamma$}
   \Text(23,28)[t]{$0$}
   \Text(55,28)[t]{$+1$}
   \Text(90,12)[t]{$+1$}
   \Text(90,57)[b]{$0$}
  \end{picture}
  \begin{picture}(120,70)(0,0)
   \Text(60,68)[t]{(b)}
   \DashLine(10,30)(40,30){5}
   \DashLine(40,30)(70,45){5}
   \Line(67,40)(73,50) \Line(67,50)(73,40)
   \DashLine(70,45)(100,60){5}
   \DashLine(40,30)(95,10){5}
   \Text(70,38)[t]{$\widetilde\Gamma$}
   \Text(23,35)[b]{$0$}
   \Text(70,12)[t]{$+1$}
   \Text(52,42)[b]{$-1$}
   \Text(90,48)[l]{$0$}
  \end{picture}

  \begin{picture}(120,70)(0,0)
   \Text(60,68)[t]{(c)}
   \DashLine(10,30)(50,30){5}
   \DashLine(30,55)(50,30){5}
   \Line(27,50)(33,60) \Line(33,50)(27,60)
   \DashLine(50,30)(100,60){5}
   \DashLine(50,30)(95,10){5}
   \Text(40,45)[bl]{$+1$}
   \Text(23,35)[b]{$0$}
   \Text(80,12)[t]{$+1$}
   \Text(90,48)[l]{$0$}
  \end{picture}
  \begin{picture}(120,70)(0,0)
   \Text(60,68)[t]{(d)}
   \DashLine(30,10)(60,30){5}
   \DashLine(30,50)(60,30){5}
   \DashLine(90,10)(60,30){5}
   \DashLine(90,50)(60,30){5}
   \Text(20,10)[c]{$0$}
   \Text(100,10)[c]{$0$}
   \Text(20,50)[c]{$\pm1$}
   \Text(100,50)[c]{$\pm1$}
   \Text(60,5)[c]{$-4f^{-2}\Delta_h$}
  \end{picture}
  \caption{\label{fig_triple_scalar_current_amplitude}
  The three diagrams (a--c) whose sum must satisfy the Ward--Takahashi 
identity. The bosonic four-point function (d) is fixed as a result.}
 \end{figure}

  Finally, we require the Ward--Takahashi identity for the sum of the 
three diagrams which are shown in 
fig.~\ref{fig_triple_scalar_current_amplitude}a--c.
  We choose the momenta to be $h(q_1)\to h(q_2)+\phi_+(q_3)$, with 
$q=q_2+q_3-q_1$ being the four-momentum flowing into the current.
 We obtain
 \begin{equation}
  q_\mu\mathcal{A}^\mu_\mathrm{a}=
  -4i^2f^{-1}\Delta_h(D_h^{-1}(q_1)D_\phi(q_1+q)+1),
 \end{equation}
  and
 \begin{equation}
  q_\mu\mathcal{A}^\mu_\mathrm{b}=
  -4i^2f^{-1}\Delta_h(D_h^{-1}(q_2)D_\phi(q_2-q)+1).
 \end{equation}
  Thus we require
 \begin{equation}
  q_\mu\mathcal{A}^\mu_\mathrm{c}=8i^2f^{-1}\Delta_h
 \end{equation}
  in order that the Ward--Takahashi identity is satisfied.
  Hence we obtain the Feynman rule shown in 
fig.~\ref{fig_triple_scalar_current_amplitude}d.

 \section{Calculation of the parameters}
 \label{sec_parameters}

  In the preceding section, we worked out the form of the theory. Let us 
now work out the parameters.

  In sec.~\ref{sec_ferro_ratio_rule}, we used the condition of tadpole 
cancellation to work out a certain rule involving the ratios of density 
of states, that need to be satisfied in ferromagnetism.

  In sec.~\ref{sec_spin_current}, we presented the Feynman rules 
and were able to relate the current--magnon two-point function to the 
form of $f^2$ and the bosonic propagators.

  We now generalize these results, and work out the four parameters, 
which are (1) $f^2$, (2) $\Delta E$, (3) $\Delta_h$ and (4) $u$ or 
$m_\phi$.

  Corresponding to these four unknowns, we have four equations, which 
involve: (1) tadpole cancellation, (2) the time component of the 
current--boson two-point function, (3) the space component of the same 
two-point function and (4) the Higgs-boson self-energy.

 \subsection{Tadpole cancellation}

  Let us start with the condition of tadpole cancellation.

  We treated the ferromagnetic case in sec.~\ref{sec_ferro_ratio_rule}. 
The anti-ferromagnetic case is calculated analogously, but the mechanism 
of cancellation is different. This time, we have the contribution of the 
magnon loop:
 \begin{equation}
  i\mathcal{A}^\mathrm{tadpole}_\mathrm{magnon}
  =i^2(2f^{-1}\Delta_h)\int\frac{d^{d+1}k}{(2\pi)^{d+1}}D_\phi(k),
  \label{eqn_magnon_tadpole}
 \end{equation}
  which must be equal and opposite to the fermionic loop:
 \begin{equation}
  i\mathcal{A}^\mathrm{tadpole}_\mathrm{fermion}
  =(-1)i^2(f^{-1}\Delta E)\int\frac{d^{d+1}k}{(2\pi)^{d+1}}
  \left(G_+(k)-G_-(k)\right).
 \end{equation}
  Note that the sum over positive and negative $\Delta E$ is implicit.

  This gives us the following condition:
 \begin{equation}
  -2\Delta_h\int\frac{d^{d+1}k}{(2\pi)^{d+1}i}D_\phi(k)=
  \sum_\mathrm{sublattices}\left[(\rho_+-\rho_-)\Delta E\right].
 \end{equation}
  Note that $\Delta E$ is positive when $\rho_->\rho_+$. That is, the 
right-hand side is negative.
  Let us introduce a more compact notation:
 \begin{equation}
  2\Delta_h\int D_\phi=\rho_M\left|\Delta E\right|.
  \label{eqn_integral_notation}
 \end{equation}
  The convention is that $\rho_M=\sum\left|\rho_-+\rho_+\right|$ is 
positive.
  The integral of $D_\phi$ is evaluated easily:
 \begin{equation}
  \int D_\phi=\int\frac{d^{d+1}k}{(2\pi)^{d+1}i}
  \frac{-1}{k_0^2-u^2\mathbf{k}^2+i0}
  =\int\frac{d^d\mathbf{k}}{2u(2\pi)^d\left|\mathbf{k}\right|}.
  \label{eqn_integral_d_phi}
 \end{equation}
  This is divergent at large momenta, and needs to be cut-off at 
$\left|\mathbf{k}\right|=K$, where $K\sim\pi/a$. We then obtain
 \begin{equation}
  \frac{\rho_M\left|\Delta E\right|}{2\Delta_h}=
  \left\{\begin{array}{ll}
   K/4\pi u & (d=2),\\
   K^2/8\pi^2u & (d=3).
  \end{array}\right.
  \label{eqn_AF_tadpole_results}
 \end{equation}
  Note that the propagators are all-order, and this requires that 
$\Delta E$ is bare, and that the vertex corrections are not included in 
eqn.~(\ref{eqn_magnon_tadpole}).
  Whether $\Delta E$ is stable against higher-order corrections depends 
on the size of the coupling $f^{-1}\Delta E$ and the relative size of 
$u$ compared with the electron velocity $v\approx v_F$.
  The form of eqn.~(\ref{eqn_magnon_tadpole}) corresponds to the all-order 
vertex, and therefore this equation, and 
eqn.~(\ref{eqn_AF_tadpole_results}) which follows from it, suffer from 
double counting.
  However, this ambiguity, that is due to double counting, cancels when 
we discuss the Higgs-boson self-energy later on.

 \subsection{Current--magnon two-point function}

  In sec.~\ref{sec_spin_current}, we calculated the fermionic 
contribution to the current--magnon two-point function, which is shown 
in fig.~\ref{fig_two_point_func_loop}a.
  We obtained
 \begin{equation}
  \mathcal{A}_\mathrm{fermionic}^\mu=
  -\int\frac{d^{d+1}k}{(2\pi)^{d+1}i}(f^{-1}\Delta E)\Gamma^\mu
  G_+(k)G_-(k-q).
 \end{equation}
  For the simple case of
$\epsilon(\mathbf{k})=(\hbar\mathbf{k})^2/2m_e$, the vertex is given by
  $\Gamma^\mu=(1,(\mathbf{k}-\mathbf{q}/2)/m_e)$.
  The bosonic loop, which corresponds to 
fig.~\ref{fig_two_point_func_loop}b, is written as
 \begin{eqnarray}
  &&\hspace{-10pt}\mathcal{A}_\mathrm{bosonic}^\mu=\nonumber\\
  &&\int\frac{d^{d+1}k}{(2\pi)^{d+1}i}(2f^{-1}\Delta_h)(-2(2k-q))
  D_h(k)D_\phi(k-q)
 \end{eqnarray}
  This is for the anti-ferromagnetic case.
  Note that the case of ferromagnetism requires a further twist, as we 
have not yet included the higgs--screened-photon mixing.

  Let us consider the limit of small external momentum, $q\to0$.
  It is easy to see that the fermionic amplitude vanishes for 
anti-ferromagnetism. As for the bosonic amplitude, this vanishes for 
ferromagnetism because of the absence of negative energy states. The 
amplitude vanishes for anti-ferromagnetism also, but for a different 
reason, namely symmetry.

  Let us, instead of trying to evaluate these integrals for arbitrary 
values of $q$, make use of the Ward--Takahashi identities to replace the 
current--magnon two-point functions with the corresponding 
current--current two-point functions (c.f., ref.~\cite{gribov}).

  To do so, we first write down the current--current two-point 
functions as
 \begin{equation}
  \Pi^{\mu\nu}_\mathrm{fermionic}
  =-\int\frac{d^{d+1}k}{(2\pi)^{d+1}i}\Gamma^\mu\Gamma^\nu
  G_+(k)G_-(k-q),
 \end{equation}
  and
 \begin{eqnarray}
  &&\hspace{-10pt}\Pi^{\mu\nu}_\mathrm{bosonic}
  =\int\frac{d^{d+1}k}{(2\pi)^{d+1}i}\nonumber\\
  &&(-2(2k-q)^\mu)(-2(q-2k)^\nu) D_h(k)D_\phi(k-q).
 \end{eqnarray}
  By virtue of the Ward--Takahashi identities, we obtain
 \begin{equation}
  q_\nu\Pi^{\mu\nu}-f\mathcal{A}^\mu=C^\mu_\mathrm{fermionic}
  +C^\mu_\mathrm{bosonic},
 \end{equation}
  where $C^\mu$ are given by
 \begin{equation}
  C^\mu_\mathrm{fermionic}=\int\frac{d^{d+1}k}{(2\pi)^{d+1}i}
  \Gamma^\mu(k,q-k)\left(G_+(k)-G_-(k-q)\right)
 \end{equation}
  and
 \begin{equation}
  C^\mu_\mathrm{bosonic}=\int\frac{d^{d+1}k}{(2\pi)^{d+1}i}
  (-2(2k-q)^\mu)\left(D_h(k)+D_\phi(k-q)\right).
 \end{equation}

  We now equate $\mathcal{A}$ with the Feynman rules of 
fig.~\ref{fig_boson_two_point_functions}c, and take the derivative with 
respect to $q_\lambda$ in the limit of small $q$. For the ferromagnetic 
case, we obtain
 \begin{equation}
  \left.\Pi^{\mu\lambda}_\mathrm{fermionic}\right|_{q\to0}+
  f^2m^{-1}_\phi\mathrm{diag}(0,-I)
  =\left.\frac{\partial}{\partial q_\lambda}C^\mu_\mathrm{fermionic}
  \right|_{q\to0}.
  \label{eqn_ferromagnetic_two_point_amplitudes}
 \end{equation}
  $I$ stands for the spatial identity matrix.
  The $0-0$ component of this equation is zero on the right-hand side 
and in the second term of the left-hand side, whereas the first term on 
the left-hand side is non-zero:
 \begin{equation}
  \left.\Pi^{00}_\mathrm{fermionic}\right|_{q\to0}=-\int G_+G_-
  =\frac{\rho_--\rho_+}{\Delta E}.
 \end{equation}
  This happens because of the approximation 
$D^{-1}_\phi(q)=q_0-\mathbf{q}^2/2m_\phi$. There is, in principle, a 
$q_0^2$ term as well, the omission of which is inconsistent with the 
$0-0$ component of this equation.
  As for the spatial components, we obtain
 \begin{eqnarray}
  &&\hspace{-10pt}
  f^2m_\phi^{-1}d=\int\frac{d^{d+1}k}{(2\pi)^{d+1}i}\nonumber\\
  &&\left[\left(G_+(k)+G_-(k)\right)\frac{d}{2m_e}+
  G_+(k)G_-(k)v_e^2\right].
 \end{eqnarray}
  Here, $d/m_e$ refers to the second derivative of 
$\epsilon(\mathrm{k})$, whereas $v_e$ refers to the first derivative.

  Let us introduce the following shorthand notation:
 \begin{equation}
  f^2m_\phi^{-1}=\left<\frac{\rho_-+\rho_+}{2m_e}-
  \frac{(\rho_--\rho_+)v_e^2}{\Delta Ed}\right>.
  \label{eqn_ferromagnetism_a}
 \end{equation}
  There is an obvious generalization to the case of spatial asymmetry.
  We expect this final result to be stable against higher-order 
corrections, because the renormalization factors due to the vertex 
correction and the Green's functions cancel.

  As discussed in sec.~\ref{sec_ferro_ratio_rule}, $(\rho_-+\rho_+)$ is 
not well-defined. However, the ratio of $(\rho_-+\rho_+)$ against 
$(\rho_--\rho_+)$ is well defined because of 
eqn.~(\ref{eqn_ferromagnetic_ratio_rule}).
  Furthermore, $f^2$ is given by $\rho_--\rho_+$.

  As an order estimation, we can say that $v_e$ can be taken to be 
almost constant near the Fermi surfaces. It would be a bad approximation 
to say that $m_e$ is also constant, but we can introduce a quantity 
$\overline{m_e}$ to be the inverse of the average inverse fermion mass.
  We then obtain
 \begin{equation}
  \frac{2\overline{m_e}}{m_\phi}\approx\frac1P-
  \frac{2\overline{m_e}v_F^2}{\Delta Ed}.
 \end{equation}
  $P$ is the spin asymmetry. $m_\phi$ is necessarily positive, but 
$\overline{m_e}$ needs not be positive although we generally expect it 
to be.
  If $\overline{m_e}$ is positive, then the inequality reads
 \begin{equation}
  P\lesssim\frac{\Delta Ed}{2\overline{m_e}v_F^2}.
 \end{equation}

  Let us now turn to the anti-ferromagnetic case. Here we need both the 
fermionic and the bosonic loops. Corresponding to 
eqn.~(\ref{eqn_ferromagnetic_two_point_amplitudes}), we now have
 \begin{equation}
  \left.\Pi^{\mu\lambda}\right|_{q\to0}+f^2\mathrm{diag}(1,-u^2I)
  =\left.\frac{\partial}{\partial q_\lambda}C^\mu\right|_{q\to0}.
  \label{eqn_antiferromagnetic_two_point_amplitudes}
 \end{equation}
  The fermionic contributions are as given above. The bosonic 
contributions are given by
 \begin{equation}
  \left.\Pi^{\mu\lambda}_\mathrm{bosonic}\right|_{q\to0}=
  -16\int k^\mu k^\lambda D_hD_\phi,
  \label{eqn_pimunubosonic}
 \end{equation}
  and
 \begin{equation}
  \frac{\partial}{\partial q_\lambda}C_\mathrm{bosonic}^\mu=
  2\mathrm{diag}(1,-u^2I)\int D_h-D_\phi,
 \end{equation}
  using the same notation as in eqn.~(\ref{eqn_integral_notation}).
  The integral over $D_\phi$ is given by eqn.~(\ref{eqn_integral_d_phi}), 
and is a positive quantity.
  The integral over $D_h$ is given by
 \begin{equation}
  \int D_h=-\int\frac{d^d\mathbf{k}}{2(2\pi)^d
  \sqrt{u^2\mathbf{k}^2+\Delta_h}},
 \end{equation}
  and this is a negative quantity.

  Altogether, we obtain
 \begin{equation}
  f^2+2\int\left(D_h-D_\phi\right)-16\int k_0^2D_hD_\phi
  =-\frac{\rho_M}{\left|\Delta E\right|},
  \label{eqn_antiferromagnetism_a}
 \end{equation}
  and
 \begin{eqnarray}
  &&\hspace{-10pt}
  f^2+2\int\left(D_h-D_\phi\right)-\frac{16}d\int\mathbf{k}^2D_hD_\phi
  \nonumber\\
  &&=\frac1{u^2}\left<\frac{\rho_-+\rho_+}{2m_e}-
  \frac{\rho_Mv_e^2}{\left|\Delta E\right|d}\right>.
  \label{eqn_antiferromagnetism_b}
 \end{eqnarray}

 \subsection{The Higgs-boson self-energy}

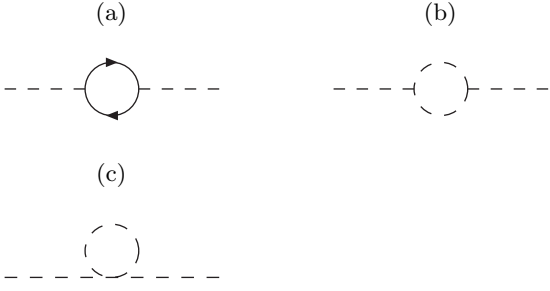
\begin{figure}[ht]
  \begin{picture}(120,60)(0,0)
   \Text(60,58)[t]{(a)}
   \DashLine(20,25)(50,25){5}
   \ArrowArcn(60,25)(10,0,180)
   \ArrowArcn(60,25)(10,180,360)
   \DashLine(70,25)(100,25){5}
  \end{picture}
  \begin{picture}(120,60)(0,0)
   \Text(60,58)[t]{(b)}
   \DashLine(20,25)(50,25){5}
   \DashCArc(60,25)(10,0,360){5}
   \DashLine(70,25)(100,25){5}
  \end{picture}

  \begin{picture}(120,60)(0,0)
   \Text(60,58)[t]{(c)}
   \DashLine(20,15)(100,15){5}
   \DashCArc(60,25)(10,0,360){5}
  \end{picture}
  \caption{\label{fig_higgs_self_energy}
  The diagrams for the self-energy of the Higgs boson.}
 \end{figure}

  We now come to the final condition, namely that the Higgs-boson 
excitation energy $\Delta_h$ is given by the self-energy diagrams which 
are shown in fig.~\ref{fig_higgs_self_energy}.

  The fermionic contribution is similar to $\Pi_z^{00}$ which was 
calculated in sec.~\ref{sec_ferro_ratio_rule}, and is given by
 \begin{eqnarray}
  &&\hspace{-10pt}-i\Pi^h_\mathrm{a}
  =-i^4\int\frac{d^{d+1}k}{(2\pi)^{d+1}}
  (f^{-1}\Delta E)^2\nonumber\\&&
  \left(G_+(k)G_+(k-q)+G_-(k)G_-(k-q)\right).
  \label{eqn_higgs_self_energy_a_derivation}
 \end{eqnarray}
  Hence
 \begin{equation}
  \Pi^h_\mathrm{a}=(f^{-1}\Delta E)^2
  \left(g_-(\epsilon_F)+g_+(\epsilon_F)\right).
  \label{eqn_higgs_self_energy_a}
 \end{equation}
  Note that $\Delta E$ refers to the bare quantity.
  This is because, firstly, $(\Delta E)^2$ in 
eqn.~(\ref{eqn_higgs_self_energy_a_derivation}) needs to be the product of 
the bare $\Delta E$ and the renormalized $\Delta E$.
  However, the renormalization of $\Delta E$ gives rise to the 
renormalization factor $Z^{-1}$ which is opposite to the renormalization 
factor $Z$ for each propagator.
  It follows, therefore, that $\Delta E$ in 
eqn.~(\ref{eqn_higgs_self_energy_a}) actually refers to the bare quantity.

  At $q=0$ (and at zero temperature), the contributions of 
fig.~\ref{fig_higgs_self_energy}b and c are zero for ferromagnetism. 
Hence, for the case of ferromagnetism, we obtain
 \begin{equation}
  \Delta_h=\frac{g_-(\epsilon_F)+g_+(\epsilon_F)}{\rho_--\rho_+}
  (\Delta E)^2.
  \label{eqn_higgs_self_energy_ferromagnetism}
 \end{equation}
  If the density of states $g$ is a linear function, then 
$\Delta_h=2\Delta E$ since $(\rho_--\rho_+)$ is given by the area of a 
trapezium whose two parallel sides are $g_-$ and $g_+$, and whose height 
is $\Delta E$. If not, and $g$ is a convex function in between $g_-$ and 
$g_+$ as is the case for iron \cite{band_calculation}, $\Delta_h$ will 
be less than $2\Delta E$. This gives a useful estimate of the 
Higgs-boson excitation energy, which can be tested experimentally.

  We should remember that the Higgs--screened-photon mixing cannot be 
neglected when the spin asymmetry $P$ is large.
  The actual value of $\Delta_h$ where the resonance occurs will be 
sensitive to the behaviour of the photonic modes (screened photon and 
plasmon).

  Let us now turn to anti-ferromagnetism.
  The contribution of fig.~\ref{fig_higgs_self_energy}b is given by
 \begin{equation}
  -i\Pi^h_\mathrm{b}=i^4\int\frac{d^{d+1}k}{(2\pi)^{d+1}}
  (2f^{-1}\Delta_h)^2D_\phi(k)D_\phi(k-q).
 \end{equation}
  This is divergent, but is imaginary at zero temperature for 
$q_0^2-u^2\mathbf{q}^2>0$ (which is where the Higgs mode needs to 
exist).
  We therefore omit this contribution for now.
  The contribution of fig.~\ref{fig_higgs_self_energy}c is given by
 \begin{equation}
  -i\Pi^h_\mathrm{c}=i^2\int\frac{d^{d+1}k}{(2\pi)^{d+1}}
  (-4f^{-2}\Delta_h)D_\phi(k).
 \end{equation}
  Now using eqn.~(\ref{eqn_integral_notation}), this reduces to
 \begin{equation}
  \Pi^h_\mathrm{c}=-4f^{-2}\Delta_h\int D_\phi(k)
  =-2f^{-2}\rho_m\left|\Delta E\right|.
 \end{equation}
  Hence,
 \begin{eqnarray}
  &&\hspace{-10pt}\Delta_h=\Pi^h_\mathrm{a}+\Pi^h_\mathrm{c}
  \nonumber\\&&=
  2f^{-2}\left|\Delta E\right|\left[
  \frac12\left(g_-(\epsilon_F)+g_+(\epsilon)\right)
  \left|\Delta E\right|-\rho_m\right].
 \end{eqnarray}
  We thus obtain the anti-ferromagnetic ratio rule:
 \begin{equation}
  \frac{g_-(\epsilon_F)+g_+(\epsilon_F)}
  {2\rho_M/\left|\Delta E\right|}>1.
  \label{eqn_antiferromagnetic_ratio_rule}
 \end{equation}
  This is satisfied if the density of states is a concave function in 
between the two Fermi energies.

  Let us denote the concavity by $\delta_c$, defined as:
 \begin{equation}
  \delta_c=\frac{g_-(\epsilon_F)+g_+(\epsilon_F)}
  {2\rho_M/\left|\Delta E\right|}-1.
  \label{eqn_concavity_definition}
 \end{equation}
  This then leads to
 \begin{equation}
  \frac{2\rho_M\left|\Delta E\right|}{\Delta_h}
  =\delta_c^{-1}f^2.
  \label{eqn_antiferromagnetism_c}
 \end{equation}
  By eqn.~(\ref{eqn_AF_tadpole_results}), we then obtain
 \begin{equation}
  f^2=\left\{\begin{array}{ll}
   \delta_cK/\pi u & (d=2),\\
   \delta_cK^2/2\pi^2u & (d=3).
  \end{array}\right.
 \end{equation}
  Small $\delta_c$ therefore leads to strong coupling $f^{-1}\Delta E$.
  We expect physically that strong coupling tends to suppress magnetism, 
because the oscillations between the two spin states will become more 
frequent.
  Our results are not affected so long as the densities of states can be 
defined.
  However, $\Delta E$ will receive a large correction through the 
electron self-energy.

 \subsection{Summary of results at zero temperature}

  Let us summarize our results.

  For the case of ferromagnetism, the parameters $\Delta E$, $f^2$,
$m_\phi$ and $\Delta_h$ are fixed by the following constraints:
 \begin{eqnarray}
  \rho_+/\rho_-&=&g_+(\epsilon_F)/g_-(\epsilon_F),
  \label{eqn_ferromagnetism_summary_a}\\
  f^2&=&\rho_--\rho_+,\label{eqn_ferromagnetism_summary_b}\\
  f^2m_\phi^{-1}&=&\left<\frac{\rho_-+\rho_+}{2m_e}-
  \frac{(\rho_--\rho_+)v_e^2}{\Delta Ed}\right>,
  \label{eqn_ferromagnetism_summary_c}\\
  f^2\Delta_h&=&\left(g_-(\epsilon_F)+g_+(\epsilon_F)\right)
  (\Delta E)^2. \label{eqn_ferromagnetism_summary_d}
 \end{eqnarray}
  Out of these equations, which are all non-perturbative, the first 
three are relations that only involve all-order quantities. In the last 
equation, $\Delta E$ refers to the bare quantity. In 
eqn.~(\ref{eqn_ferromagnetism_summary_c}), $\Delta E$ is the all-order
quantity, but is assumed to be more or less independent of energy and 
momenta (though generalization is possible).

  For the case of anti-ferromagnetism, $\left|\Delta E\right|$, $f^2$, 
$u$ and $\Delta_h$ are fixed by eqns.~(\ref{eqn_AF_tadpole_results}), 
(\ref{eqn_antiferromagnetism_a}), (\ref{eqn_antiferromagnetism_b}) and 
(\ref{eqn_antiferromagnetism_c}).
  Equations (\ref{eqn_AF_tadpole_results}) and 
(\ref{eqn_antiferromagnetism_c}) involve $\left|\Delta E\right|$ as a bare 
quantity and 
$\Delta_h$ in eqn.~(\ref{eqn_antiferromagnetism_c}) is ambiguous.
  Equations (\ref{eqn_antiferromagnetism_a}) and 
(\ref{eqn_antiferromagnetism_b}) only involve all-order quantities, but 
are dependent on the UV cut-off, as is the case in 
eqn.~(\ref{eqn_AF_tadpole_results}).

  We obtained the rule $\delta_c>0$, where $\delta_c$ is a measure of 
concavity and is defined by eqn.~(\ref{eqn_concavity_definition}).

 \subsection{Finite temperature analysis}

  Since our results involve diagrams that are evaluated for $q=0$, it 
is, in principle, straightforward to generalize them to finite 
temperatures.
  However, the bosonic diagrams, which were zero in the case of 
ferromagnetism, become non-zero at finite temperatures, and therefore 
the resulting expressions are messy.

  A full calculation is beyond the scope of this present analysis, but 
let us present two representative results.

  First, for the case of anti-ferromagnetism, we have found that the 
bosonic loop of fig.~\ref{fig_higgs_self_energy}b is real and diverges 
for finite $T$:
 \begin{equation}
  \Pi^h_{\mathrm{b}}(T)=
  -\frac{(2f^{-1}\Delta_h)^2}{32T^3}
  \int\frac{d^d\mathbf{k}}{(2\pi)^d}\left.
  \frac{d}{dx}\left(-\frac{\mathrm{coth}(x)}{x}\right)
  \right|_{x=u\left|\mathbf{k}\right|/2T}.
 \end{equation}
  This makes $\Delta_h$ negative, and so magnetic order is forbidden.
  In our opinion, this implies that in anti-ferromagnetic metals, a 
genuine long-range order is not permitted, at least at finite 
temperatures.

  Second, let us consider how the ratio rule of 
eqn.~(\ref{eqn_ferromagnetic_ratio_rule}) is modified at finite 
temperatures. We now have the bosonic contribution which reads
 \begin{equation}
  \mathcal{A}^\mathrm{tadpole}_\mathrm{boson}(T)=
  -(2f^{-1}\Delta_h)\int\frac{d^d\mathbf{k}}{(2\pi)^d}
  \frac1{\exp((\mathbf{k}^2/2m_\phi)/T)-1}.
 \end{equation}
  This is evaluated easily using standard methods. For the case of three 
spatial dimensions, we obtain
 \begin{equation}
  \mathcal{A}^\mathrm{tadpole}_\mathrm{boson}(T)=
  -2f^{-1}\Delta_h\zeta(3/2)\left(\frac{m_\phi T}{2\pi}\right)^{3/2}.
 \end{equation}
  Here, $\zeta(3/2)=2.612\cdots$.
  This contribution should be equal and opposite to the fermionic 
contributions, which are given by
 \begin{equation}
  \mathcal{A}^\mathrm{tadpole}_\mathrm{fermion}(T)=
  (f^{-1}\Delta E)\left[-(\rho_\downarrow-\rho_\uparrow)_T
  +(\rho_\downarrow+\rho_\uparrow)_TP(T)\right].
 \end{equation}
  Here $\rho$ and $P$ correspond to their finite-temperature 
counterparts:
 \begin{eqnarray}
  \rho_T&=&\int f((\epsilon-\mu)/T)g(\epsilon)d\epsilon,\\
  g_T&=&-\int f'((\epsilon-\mu)/T)g(\epsilon)d\epsilon,
 \end{eqnarray}
 where $f$ is the Fermi distribution function.
 Hence
 \begin{equation}
  -(\rho_\downarrow-\rho_\uparrow)_T
  +(\rho_\downarrow+\rho_\uparrow)_TP(T)=
  2\zeta(3/2)
  \left(\frac{m_\phi T}{2\pi}\right)^{3/2}\frac{\Delta_h}{\Delta E}.
 \end{equation}
  $\Delta_h$, $m_\phi$ and $\Delta E$ are also functions of temperature.

  For small $T$, we can assume that only the first term on the left-hand 
side depends significantly on $T$, and that the parameters on the 
right-hand side can be taken as constants. This then implies that the 
magnetization goes down as $T^{3/2}$, which is a well-known result.
  All of the parameters on the right-hand side are, in principle, 
measurable. This can then be tested experimentally.

 \section{Conclusions and Outlook}
 \label{sec_conclusions}

  We presented a nonperturbative framework for treating magnetic order 
in metals, caused by a Coulomb interaction (or generalized Coulomb 
interaction).

  We obtained interesting `ratio rules' involving the densities of 
states for both ferromagnetic and anti-ferromagnetic cases.
  These involve all-order quantities (with the exception of 
$\left|\Delta E\right|$) and can therefore be compared directly with the 
experimental numbers, if they become available at greater precision.

  We have seen that the shape of the density-of-states curve play an 
essential role in determining the possibility of magnetic ordering. The 
density of states must rise with energy, when the charge carriers are 
electrons, for ferromagnetism.

  For anti-ferromagnetism, the density-of-states curve must be concave.
  However, we have seen that the radiative corrections, due to the 
magnons, at finite temperatures breaks long-range orders. In our 
understanding, this means that genuine long-range anti-ferromagnetic 
order is not possible, at least at finite temperatures. More work is 
required to elucidate the nature of the ground state.

  The case of magnetic insulators is not covered by this work, in which 
the exchange energy $\Delta E$ is considered to be more or less 
independent of $\mathbf{k}$.

  Two cases require special attention, which we have not been able to 
discuss in much detail.
  The first is the case of strong coupling, which occurs when $f$, or 
the vacuum-expectation value $v$ of the magnetic order-parameter field, 
is small.
  Here, we expect that the radiative corrections suppress the magnetic 
order and that the system will favour the paramagnetic state.
  The second is the case of large magnon velocity $u$, in comparison 
with the electron velocity $v_F$, in the case of anti-ferromagnetism. 
  Here, the response of the magnetic background becomes instantaneous 
towards the movement of the electron.
  We hope to be able to discuss this case in a separate publication
\cite{odagiri_gribovequation_paper}

  The results of this work can be used to calculate arbitrary 
amplitudes, such as scattering amplitudes.

  The methods presented in this work, being an adaptation of Gribov's 
analysis of axial-current conservation, is of a general nature.
  However, we are presently unaware of other possible applications of 
the methods presented herein.

 \begin{acknowledgement}

  We thank I.~Hase, S.~Sharma, K.~Yamaji and T.~Yanagisawa for extensive 
and informative comments and discussions.

  We have been informed by Dr.~I.~Hase that a phenomenological study of 
the correlation between densities of states of a material and its 
magnetic properties has previously been reported.
  However, we have not been able to locate this study.

 \end{acknowledgement}


\begin{thebibliography}{MM}

 \bibitem{leutwyler} See, for example:
 H.~Neuberger and T.~Ziman, \jouprb{39}{1989}{2608};
 H.~Leutwyler, \jouprd{49}{1994}{3033}.

 \bibitem{gribov} V.~N.~Gribov,
\epjc{10}{1999}{71}
[arXiv:\hepph{9807224}];
\ibid{10}{1999}{91}
[arXiv:\hepph{9902279}].

 \bibitem{gribovewsb} V.~N.~Gribov, \plb{336}{1994}{243}
 [arXiv:\hepph{9407269}].

 \bibitem{gribovlectures} V.N.~Gribov,
\textit{Orsay lectures on confinement},
arXiv:\hepph{9403218}, arXiv:\hepph{9407269}, arXiv:\hepph{9905285}.

 \bibitem{yurireview} For a review, see: Yu.L.~Dokshitzer and 
D.E.~Kharzeev,
\hepph{0404216}.

 \bibitem{njl} Y.~Nambu and G.~Jona-Lasinio, \joupr{122}{1961}{345}; 
\joupr{124}{1961}{246}.

 \bibitem{topcondensation} G.~Cveti\v{c}, \jourmp{71}{1999}{513}.

 \bibitem{electrongas} See, for example:
 D.~P.~Young, et al., \nature{397}{1999}{412}, and references therein.

 \bibitem{spin_asymmetry_A} P.~M.~Tedrow and R.~Meservey, 
\prep{238}{1994}{173}.

 \bibitem{spin_asymmetry_B} R.~J.~Soulen Jr., et al.,
\newjournal{Science}{}{282}{1998}{85}.

 \bibitem{band_calculation} R.~Maglic, \jouprl{31}{1973}{546}; 
K.~C.~Wong, E.~P.~Wohlfarth and D.~M.~Hum, \pla{29}{1969}{452};
J.~Callaway and C.~S.~Wang, \jouprb{7}{1973}{1096}.

 \bibitem{Kittel} C.~Kittel, \textit{Introduction to solid state 
physics}, 5th 
edition, John Wiley \& Sons, Inc., New York, London, Sydney, Toronto, 
1976.

 \bibitem{Hall_effect} D.~E.~Gray (contributing editor), \textit{AIP 
Handbook}, 3rd edition, AIP, New York, 1972.

 \bibitem{odagiri_gribovequation_paper} See sec.~III of K.~Odagiri and 
T.~Yanagisawa, arXiv:1104.1247v1. This publication is currently under 
major revision.

 \end{thebibliography}
 \end{document}